\documentclass[11pt,a4paper]{article}

\usepackage{SGpreprint} 
\usepackage{graphicx} 
\usepackage{subfigure}
\usepackage{amsmath}

\nologohead{ 
 Patrick Groeber, Frank
Schweitzer, Kerstin Press: \\ \emph{How groups can foster consensus - The
case of local cultures} \\ Journal of Artificial Societes and Social
Simulationsn vol 12, no 2/4 (2009) \\ See \url{http://www.sg.ethz.ch} for more
information.
}

\begin{document} 

\renewcommand{\epsilon}{\varepsilon}

\begin{center}

   \textbf{\Large How groups can foster consensus: \\ The case of local cultures}\\[5mm]
   \textbf{\large Patrick Groeber, Frank Schweitzer, Kerstin Press}

       Chair of Systems Design, ETH Zurich, Kreuzplatz 5, 
       8032 Zurich, Switzerland \\
         \texttt{\{pgroeber,fschweitzer,kpress\}@ethz.ch}
\end{center}

\begin{abstract}
  A local culture denotes a commonly shared behaviour within a
  cluster of firms. Similar to social norms or conventions, it is an emergent
  feature resulting from the firms' interaction in an economic
  network. 
  To model these dynamics, we consider a distributed agent population,
  representing e.g. firms or individuals. Further, we build on a continuous
  opinion dynamics model with bounded confidence ($\epsilon$), which assumes
  that two agents only interact if differences in their behaviour are
  less than $\epsilon$. Interaction results in more similarity of 
  behaviour, i.e. convergence towards a common mean. This framework is
  extended by two major concepts: (i) The agent's in-group consisting of
  acquainted interaction partners is explicitly taken into account. This
  leads to an effective agent behaviour reflecting that agents try to
  continue to interact with past partners and thus to keep sufficiently
  close to them. (ii) The in-group network structure changes over time,
  as agents can form new links to other agents with sufficiently close
  effective behaviour or delete links to agents no longer close in
  behaviour. Thus, our model provides a feedback mechanism between the
  agents' behaviour and their in-group structure. Studying its
  consequences by means of agent-based computer simulations, we find that
  for narrow-minded agents (low $\epsilon$) the additional feedback helps
  to find consensus more often, whereas for open-minded agents (high
  $\epsilon$) this does not hold. This counterintuitive result is
  explained by simulations of the network evolution.

\end{abstract}

\begin{center}
\end{center}

\section{Introduction}
\label{intro}

Since the early $1990$s, economic theory and policy has dedicated a lot
of attention to the causes and effects of clusters, i.e. spatial
concentrations of firms in one or a few related industries
\citep{ellison.ea97}.  Driven by their impressive prosperity, the factors
underlying cluster success were extensively studied in the hope of
replicating areas like the Silicon Valley or London.  Locating in a
cluster provides several benefits for firms.  However, the existence of these
benefits hinges on a set of rules
on acceptable business practice -- the cluster's \emph{`local culture'}.  For
instance, there is a rule among banks located in Frankfurt that they should not
hire personnel away from competitors.  In the Silicon Valley, the local culture
is such that firms openly exchange ideas even with direct competitors.  In
order to partake in the cluster and its dynamics, firms need to respect these
rules, which has made it difficult for outside companies (e.g. multinational
enterprises) to tap into the cluster through subsidiaries 
 While some work has been conducted to determine the
nature and enforcement of such local cultures, their emergence remains far
from understood.  So far, anecdotal evidence suggests that it
depends on agreement about the nature of desirable business practice.

To shed light on the emergence of local cultures, the present paper
studies the emergence of the initial consensus.  It argues that cluster firms
have to interact with each other.
As these interactions are not cost-free, they give rise to
changes in
firm behaviour and different inter-firm networks that affect the future
behaviour of constituent firms.  Both mechanisms can lead to
convergent or divergent behaviour.  Those cases finding converging
behaviour among all or a majority of cluster firms constitute situations
in which the basis for a local culture emerges.  This is also
important for the emergence of clustering benefits and therefore the
economic viability of the cluster.  To study the emergence of converging
behaviour (i.e. the first step towards a local culture), this paper develops an 
opinion dynamics model with bounded confidence and group-influence as both
aspects are required to mimic the dynamics of inter-firm interaction in
clusters.

In doing so, the paper proceeds as follows.  Section \ref{sec:whylc}
describes the case under study.  The emergence and effect of local
cultures is found to be similar to that of norms and conventions as both are
means to solve co-ordination and cooperation problems.
However, the costliness of inter-firm interactions requires some
amendments to existing models (section
\ref{sec:localcultures}).  In particular, it justifies the application of
a bounded confidence model introduced in section \ref{sec:model}:  Agents (=
firms) will only interact if they are sufficiently
close in their behaviour as the risk of a costly transaction going wrong are
too high otherwise.  Moreover, agents are linked by ties stemming from
previous interactions.  Thanks to costly interaction, agents try to
maintain these `in-groups' and modify their behaviour accordingly.
Section \ref{sec:results} then presents the findings regarding the
conditions for consensual behaviour among all or a majority of agents.  It is
found that the existence and influence of in-groups fosters consensus,
especially if agents would only interact with behaviourally similar actors.
As a result, the nature of inter-firm interactions in clusters is conducive to
consensus and thereby the emergence of a local culture.

\section{Defining local cultures}
\label{sec:whylc}

In economic geography, local cultures with rules like "do not hire from
competitors", "exchange ideas freely" or "deliver only the highest quality" are
a phenomenon characterising
clusters like Silicon Valley (IT), London (financial services) or Prato,
Italy (textiles).  The existence of clusters is tied to that of local
cultures because the benefits to clustering are subject to various
co-ordination and co-operation issues that are solved by the rules in the
local culture.  

Usually, cluster benefits relate to scale and
specialisation effects as well as positive externalities.  The former
(scale and specialisation) emerge since companies in a cluster usually
divide the production process.  Rather than having all firms manufacture
shoes, one specialises in soles while other provide laces, tops, linings
and so on \citep{pyke.ea90}.  This division of labour leads to
\textit{scale and specialisation} benefits as firms can achieve a greater
output with a limited budget and become more efficient in their
activities \citep{smith03}.  Second, companies conduct competing and
complementary activities under identical local conditions in the cluster.
This leads to \textit{positive externalities} in the diffusion of ideas
and the availability of skilled labour.\footnote{Competitors experiment
with different strategies under identical conditions.  This allows for
direct comparisons of performance and selection of best practice.  In
addition, firms tackling the same or related problems may exchange
knowledge through various mechanisms \citep{allen83}.  Both aspects
contribute to \textit{knowledge spillovers}.  Finally, many firms in a
cluster increase the quality of the local \textit{labour pool} by
training activity and immigration of skilled people.}

To generate these benefits, cluster firms have to overcome several
dilemmas.  For a division of labour to emerge, suppliers need a fair price
and a sufficient market \citep[p. 27]{smith03}.  Moreover, the quantities
provided by different firms have to be aligned.  
Positive externalities in knowledge and personnel are tied to respective
investments in research and training.  This is only viable if
defection (free-riding) is limited.\footnote{Investment research and training
requires an understanding that allows firms to capitalise on it, i.e. no
exploitation of others' efforts or hiring away of personnel trained elsewhere.} 
Akin to famous cooperation problems like
the Kula Ring \citep{ziegler07} or the Chicago diamond market
\citep{coleman90}, the cluster literature argues, that these dilemmas are
solved by rules on acceptable business practice that make up the
\textit{local culture}.  While some work has determined what local
cultures look like (e.g. \citealt{porter90,pyke.ea90,saxenian94}) and how
their enforcement can be ensured \citep{hollaender90,kandel.ea92} their
\textit{emergence} is far less understood.  

It is usually suggested that
firms learn about successful behaviour when interacting with others.
Successful past behaviour is then repeated and possibly copied among firms in
the cluster, which implies that gradually, this behaviour spreads in the
population.  Once etablished, such a consensus on "good" behaviour creates
expectation about others' future behaviour, thereby reducing frictions in
interaction.  Specific rules making up a local culture can finally emerge to
foster and enforce this consensual behaviour by monitoring and punishment of
defecting agents \citep[p. 926]{maskell01}.
The first step towards a local culture is thus a behaviour that is viewed
as desirable enough to become the basis of rule-making.  We argue that a
good candidate for viable rules is a behaviour already prevalent in the
cluster, i.e. a behaviour that is shared by (a majority of) local firms.
As a result, \textit{consensus} on a certain behaviour constitutes the
necessary condition for rule-making, rule enforcing and the emergence of
local cultures.  This paper investigates how a consensus on a specific
behaviour emerges in a cluster.  In doing so, the paper argues that the
division of labour in clusters requires interaction between firms to
manufacture the product.  As interactions are not cost free
\citep{coase1937,williamson75}, two mechanisms come into play.

First, interaction costs increase with the difference in
firm behaviour.  Firms with similar behaviour are likely to respond
similarly to future developments.  This makes it unnecessary to specify
all possible scenarios in a contract thereby reducing the cost of
interaction.  If all interactions are equally beneficial, the increasing
cost of interaction implies that (a) firms will not interact with all
possible partners and (b) the partners to an interaction modify their
behaviour to become more similar.  Second, costly interactions make it
beneficial to maintain links with existing partners.  Firms therefore
become more embedded in networks emerging from their interaction history.
Moreover, the desire to maintain these networks may constrain their
behaviour as firms seek to remain sufficiently similar to existing
partners.

The consensus resulting from these mechanisms (if any) can take very
different forms.  It can reside with a behaviour that is very
co-operative, i.e. every firm strongly invests in activities subject to
externalities and supplier-buyer relations are characterised by fairness.
Such a situation provides high incentives to free-ride implying that this
behaviour is not self-enforcing.  In other instances, consensus can
reside with very defective behaviour where all firms try to exploit one
another as much as possible and do not investment in activities with
externalities.  This situation would provide no incentive for deviation
(at least not to an individual firm), i.e. the local culture is
self-enforcing.

\section{Local cultures as norms or conventions}
\label{sec:localcultures}

Once established, local cultures fulfill the function of social norms or
conventions insofar as they solve co-operation or co-ordination
problems. As a result, their emergence mimics that of norms and conventions,
which unfolds as follows:
The first stage is build on consensus formation, where agents reach consensus
on a certain behaviour through different mechanisms like optimising behaviour
\citep{weber99,opp82}, imitating or replicating successful strategies
  \citep{asch56,sherif73} or through trial and error search
  \citep{demsetz67}.
Once the consensual behaviour spreads and remains in the population, it
creates expectations about everyone's future behaviour, which reduce
friction in interactions \citep{axelrod86,koford.ea91,sugden89}.
Depending on whether the behaviour is self-enforcing (convention) or not
(norm), the second stage of the process differs.  For
conventions, the emergence of consensus is sufficient.  In case of norms
behavioural regularities have to result in a sense of ``oughtness''
\citep{opp01} that may eventually lead to an enforceable norm prescribing
this behaviour: ``\textit{Thus, patterns of action emerge that then
  become normative [...].  Individuals comply with the new norm both for
  the original reason that the behaviour was appealing, and also because
  it is now socially enforced}'' \citep[p. 6]{horne01a}.

Depending on the nature of consensual behaviour (self-enforcing or not),
local cultures correspond to conventions or social norms.  In either case, the
first stage of their emergence process (consensus building) is identical.  This
makes models on the emergence of norms or convention applicable to our case.
In the literature, most work on norms and conventions is based on game-theory
with a smaller subset of research studying consensus formation (through voter
models or bounded confidence approaches).  The game-theoretic approach to norms
and conventions
\citep{fehr.ea04,voss01,hollaender90,kandel.ea92,UllmannMargalit77} is not
applicable to our case for several reasons. By focussing on the nature of the
game (e.g. payoff structure, repetition) and underlying agent interaction,
individual incentives and efficient outcomes are derived.  Conventions
correspond to situations
where the optimal outcome is a non-unique Nash equilibrium.  Depending on
the underlying mechanisms, different equilibria may be selected
\citep[see e.g.][]{shoham92,
pujol05}. Norms
instead emerge in situations where the optimal solution is not an
equilibrium outcome.  The norm is viewed as the solution to the problems
preventing a better outcome in that particular game.\footnote{Candidate
mechanisms are repeated games, reputation formation, signalling
mechanisms or punishment \citep{nowak06}.  To be effective, consensus
about the behaviour that \textit{ought} to be adopted is needed
(e.g. cooperation).  To some, this would already constitute a norm
\citep{opp01}.  In addition to strong information requirements,
game-theoretic approaches thus need a general co-operation `norm' to
identify defectors.}

The focus on payoffs and incentives implies that there
\textit{is} a known optimal behaviour. In other words, agents know the
consequences of their actions, anticipate the choices of others and are
thereby able to determine, what kind of behaviour leads to efficient
outcomes.  In the Prisoner's Dilemma situation, cooperation is the
ex-ante optimal behaviour when jointly maximising the players' outcome.
Moreover, game theory is less concerned with the emergence of a particular norm
or convention but rather with its effect for the game's equilibrium outcomes. 
Since we cannot determine ex-ante payoff values for business strategy and our 
concern resides more with the emergence than the effect of local cultures,
game-theoretic models are not suited for our research question. 
We therefore focus on an approach that does not assume
an `optimal' behaviour but where the value of an agent's behaviour
only depends on the number of agents adhering to it and on that behaviour's
compliance with the predominant behaviour in the agent's personal
network. In this sense, any consensus among all agents is `good' -
regardless of the nature of consensual behaviour.

There are several models studying the emergence of consensus from agent
interaction \citep{axelrod97,lehrer.ea81,degroot74,deffuant00,hegselmann.ea02}. 
They fall into two main classes: Voter models and bounded confidence models.
In voter models, agents are characterised by a \emph{discrete opinion} (a binary
variable in most cases) and are embedded in a network of given topology. They
may adopt other opinions according to their frequency in the agent's 
neighborhood. In linear voter models, the transition towards a given
opinion is directly proportional its local frequency.  In non-linear voter
models other types of frequency dependent behaviour are
possible \citep{fs-wcss-06}. While consensus is always reached in linear voter
models, non-linear responses to the local frequency of an opinion may prevent
\citep{behera-fs-08} or accelerate \citep{stark-prl-08,stark-acs-08}
consensus.\footnote{\citet{stark-prl-08,stark-acs-08} have discussed a
modification of the linear voter model, where agents become more
reluctant to change their current opinion the longer they have it. It was
shown that this deceleration of the individual dynamics may, for certain
growth rates of the reluctance, even accelerate the formation of
consensus on the system's level. This counterintuitive result is partly
due to the agent's \emph{heterogeneity}, i.e. differences in their
individual behaviour.}

Another class of consensus models deals with \emph{continuous opinions}
$x_i$ represented as a real number between $0$ and $1$
\citep{deffuant00}. Two agents $i$ and $j$, randomly chosen at each
timestep, can only interact if the difference in their opinion does not
exceed a threshold value $\varepsilon$.  Rather than taking place on a
predifined network, agent interactions are randomised and conditional here. 
This mechanism of `bounded
confidence' is applied by \citet{hegselmann.ea02} where all
possible interactions take place simultaneously. As
investigated by means of several approaches \citep[e.g.][]{lorenz06,ben-naim03}
consensus then largely depends on the value of the key parameter
$\epsilon$.\footnote{\citet{deffuant05} incorporate an extension of
the bounded confidence mechanism in a model of innovation diffusion. For a
survey of results on bounded confidence models see \citet{lorenz07}.}

Our model, formalized in the following section, builds on the bounded
confidence approach, but combines it with the consideration of the dynamics in
an agent's
\emph{social network}.
Dependent on the relationship
between $i$ and $j$, we distinguish between an \emph{in-group}
(members of a social network seen as friends) and an \emph{out-group}
(members with adversary relations, enemies as in
\citet{norms-acs-07}).\footnote{\citet{norms-acs-07} have
investigated the role of in- and out-group interaction (in a fixed
network) for the emergence of social norms. Under certain
conditions, convergence towards a single norm, coexistence of two opposing
norms, and coexistence of a multitude of norms can be found. The key
element in this dynamics was a local utility maximization, where agent
$i$ tends to maximise similarities with her in-group, while minimising
the disutilities from the out-group in her social network.}
In addition to in-group relations within a
social network, we include a \emph{dynamics} of the agents'
social network. This is based on a feedback mechanism between an agent's
behaviour and her personal network: past interactions with partners from
the agent's in-group affect her individual behaviour which in turn influences
the structure of the in-group, iteratively. Hence, as the
novel element, our model combines both \emph{opinion dynamics} and \emph{network
dynamics} at the level of individual agents.

Both aspects relate to the fact that interaction between agents is not
cost-free.  First, we argued that costs increase with differences in
agent behaviour since many actions are beneficial as long as both agents
behave in the same way.\footnote{The greatest losses often result from diverging
  strategies (e.g. prisoner's dilemma).}  As a consequence, interactions
are conditional on sufficient behavioural similarity.  Moreover,
interacting agents approach each other's behaviour to lower interaction
cost.  Second, costly interactions make keeping past partners very
beneficial.  As a result, agents will want to keep their past partners
(their in-group) and will modify their behaviour accordingly.  The
behaviour of the group will therefore determine the possible range of
agent behaviour. Conversely, agent behaviour feeds back on her in-group
and therefore allows for changes in group structure as a function of interaction
and behavioural dynamics.  Section \ref{sec:model} provides more detail on the
formal treatment of these mechanisms.

\section{Modelling the emergence of consensual behaviour}
\label{sec:model}

We argued before that costly interactions result in
two effects, (i) \emph{consensus formation}, i.e. optimising behaviour
within a group to avoid friction, and (ii) \emph{network formation},
i.e. optimising the agents' social network structure by deleting links
with agents whose behaviour largely deviates from one's own, thus making 
interaction more costly (or creating links with agents whose behaviour is more
similar). The two interlinked dynamics are specified as follows.

\textbf{Consensus formation.} In order to reflect the first consequence
of costly interactions, we need a model that makes interaction
conditional on agent behaviour. Following \citet{deffuant00}, agent $i$'s
behaviour $x_i$ is represented as a real number between $0$ and
$1$. Thus, we are able to measure the distance between two agents
behaviour and to model a gradual approach if these agents interact. This
is different to the cultural dissemination framework of \citet{axelrod97}
where cultures constitute a finite, discrete and in general non-metric
set. There, interaction between two agents can only lead to complete
assimilation of behaviour for one of them, whereas agents approach each other's
behaviour in our model. We further define
the behaviour profile $x=(x_1,...,x_n)$.

In our model, in accordance with \citet{deffuant00}, two agents 
$i$ and $j$ are randomly chosen at each timestep. They can only interact
if the difference in their behaviour does not exceed a threshold value
$\varepsilon$ which can be regarded as a measure of \emph{openness}. 
Regarding agent interaction, we can also interpret
$\varepsilon$ as the difference in behaviour where the costs and benefits
break even:  With greater behavioural differences,
interaction costs would increase while the benefits are assumed
to be constant.  Such an interaction would lead to a net cost to the
agents involved and will therefore not occur.\footnote{As in
  \citet{moscovici.ea92} where opinions too far from the majority don't
  enter group discussion.}  As the benefits and costs are identical for
all agents, the necessary condition for an interaction of $i$ and $j$
becomes:
\begin{equation}
\label{eq:firststep} |x_i(t)-x_j(t)|<\varepsilon.
\end{equation}

If two agents interact, they try to maximise the benefits of this
exchange.  As benefits are constant and costs decrease with behavioural
differences, the behaviour of interacting agents becomes more similar as
both approach each other by identical amounts: 
\begin{equation}
\label{eq:approach}
\begin{array}{ll} x_i(t+1) = x_i(t) + \mu(x_j(t)-x_i(t)) \\ x_j(t+1) =
x_j(t) + \mu(x_i(t)-x_j(t)).
\end{array}
\end{equation} 
The speed of approach in behaviour depends on the parameter $\mu$.  It
reflects the well established phenomenon that interacting parties become
more similar
(e.g. \citealt{axelrod97,macy.ea98,strang.ea98,mcpherson.ea01}).

The dynamics specified by Eqs. (\ref{eq:firststep}), (\ref{eq:approach})
are referred to as the \emph{baseline model} in the following, as they
result in the known behaviour already discussed by \citet{deffuant00}.  We
now extend this model by introducing the second aspect: Costly
interactions imply benefits to keeping past partners.  This is modelled
by aggregating each agent's past interaction partners.  Each agent $i$
thus has a set $I_i$ of other agents constituting her \textit{in-group}, i.e.
the agent's acquainted partners.  As the agent would like to
interact with these partners later, she tries to keep her behaviour
sufficiently similar to them.  As agent behaviour changes with her
interactions, we argue that the in-group exerts an influence on the
agent's future interactions.  This is achieved by combining an agent's
behaviour $x_i$ and the mean behaviour of her in-group $\bar{x}_{I_i(t)}$ to
determine the \textit{effective behaviour}
\begin{equation}
\label{eq:effbehaviour} x_i^{\textrm{eff}}(t) =
\big(1-\alpha_i(t)\big)x_i(t) + \alpha_i(t) \bar{x}_{I_i(t)}
\end{equation}
at time $t$. The use of the mean behaviour is chosen to mirror that the
agent is equally interested in interacting with any of her past
partners.\footnote{This treatment of group influence by averaging has a
  long tradition.  Formal models of group decision-making
  \citep{french56,harary59,lehrer75,lehrer.ea81,wagner78,hegselmann.ea02}
account for
  group influence by weighted averages. Social impact theory
\citep{latane81,latane.ea97} also
  constructs group influence by averaging.  Similar to social impact
  theory, our model features a decreasing marginal group influence with
respect to adding
  more agents to the in-group.}

In Eq. (\ref{eq:effbehaviour}), $\alpha_i \in [0,1]$ corresponds to the
influence of agent $i$'s in-group on her effective behaviour.  We use
this parameter to mirror the strength of group influence. Based on the
aforementioned notion that agents like to keep their past partners, the
influence of the group would increase with its size.  In this model, we
define $\alpha_i$ endogenously by
\begin{equation}
\label{eq:group} \alpha_i(t) = \frac{|I_i(t)|}{|I_i(t)|+1}.
\end{equation}
Hence, we assume that each agent puts equal weight on her own and each
in-group member's behaviour. If $i$ has
never interacted with another agent, her in-group is empty
($|I_i|=0$) implying $\alpha_i=0$. Hence, the
effective behaviour of agent $i$ is then identical to her behaviour
($x_i^{\textrm{eff}}=x_i$).  Further, $\alpha_i$ approaches $1$ with
growing in-group size $|I_i|$.  If the in-group is large, $i$'s effective
behaviour will therefore tend towards the average in-group behaviour.

In our model, two agents wanting to interact now have to compare the distance
between their effective behaviour (influenced by their respective
in-groups) instead of that of their own behaviour.  As a result, the
necessary
condition for interaction between agents becomes
\begin{equation}
\label{eq:interaction}
|x_{i}^{\textrm{eff}}(t)-x_{j}^{\textrm{eff}}(t)|<\varepsilon.
\end{equation}
If two agents with empty in-groups interact, this is identical
to Eq. (\ref{eq:firststep}) as $x_i^{\textrm{eff}}=x_i$ and
$x_j^{\textrm{eff}}=x_j$ for $I_i=I_j=\emptyset$.

\textbf{Network formation.} In addition to maintaining their in-group, agents
also seek to expand it with suitable new partners. To specify this, we assume
that in-groups are initially empty for all agents. Later, they evolve
according to the agents' interactions as follows: In each simulation
step, two agents $i$ and $j$ are randomly selected.  If
Eq. (\ref{eq:interaction}) holds for them, they interact and are
added to each other's in-group (if they are not already contained).  Over
time, agents $i$ and $j$ may interact with different agents.  Therefore,
their effective behaviour can be altered either directly due to a change
of $x_i$ and $x_j$ resulting from interaction with other agents, or indirectly
by interactions of agents in their in-group affecting the average behaviour of
the respective in-group. Thus, we may
encounter a situation where agents $i$ and $j$ interacted at time $t$ and
were added to each other's in-group while later at $t'>t$, their
effective behaviour may be modified such that
$|x_{i}^{\textrm{eff}}(t')-x_{j}^{\textrm{eff}}(t')|\geq\varepsilon$. In this
case, agents $i$ and $j$ could no longer interact when selected and would be
removed from each other's in-group.

Thus, if $i$ and $j$ are selected at time $t$, we have
\begin{equation}
\label{eq:group_evolution}
\begin{array}{l} I_i(t+1) =
\begin{cases} I_i(t) \cup \{j\} & \text{ if }
|x_{i}^{\textrm{eff}}(t)-x_{j}^{\textrm{eff}}(t)|<\varepsilon \\ I_i(t)
\,\backslash\, \{j\} & \text{ if }
|x_{i}^{\textrm{eff}}(t)-x_{j}^{\textrm{eff}}(t)|\geq\varepsilon
\end{cases}, \\ I_j(t+1) =
\begin{cases} I_j(t) \cup \{i\} & \text{ if }
|x_{i}^{\textrm{eff}}(t)-x_{j}^{\textrm{eff}}(t)|<\varepsilon \\ I_j(t)
\,\backslash\, \{i\} & \text{ if }
|x_{i}^{\textrm{eff}}(t)-x_{j}^{\textrm{eff}}(t)|\geq\varepsilon
\end{cases}.
\end{array}
\end{equation}
Note that the in-group relation is symmetric but may not be transitive,
i.e. agent $i$ being contained in agent $j$'s in-group and agent $j$
being contained in agent $k$'s in-groups does not require $k$ being in $j$'s
in-group.

As indicated before, the behaviour of interacting agents becomes more
similar. This means that interaction at time $t$ alters the agents'
behaviour according to Eq. (\ref{eq:approach}).  For $t+1$, this also
feeds back on the effective behaviour of $i$ and $j$,
Eq. (\ref{eq:effbehaviour}), as well as on the effective behaviour of
agents whose in-group contains $i$ or $j$. 
{The effect of modifying $i$'s behaviour for her effective behaviour will
  decrease with larger $I_i$.}  Over time, the agent can interact with
others in and outside her in-group if Eq. (\ref{eq:interaction}) is
satisfied. This influences her behaviour as well as the evolution of her
in-group over time.  Agents previously outside $i$'s in-group are added
to the set $I_i$ once $i$ successfully interacts with them. The addition
of new agents to $I_i$ influences her effective behaviour and thereby her
potential for future interaction.\footnote{Many models in sociology build
  upon a reciprocal link of behaviour and interaction.  See
  \citet{carley91,coleman61,coleman80,friedkin.ea90,marsden.ea93,nowak.ea90}. 
  Empirically, the mechanism has been observed by \citet{ennett.ea94} and many
  others.}  Thus this model provides a \emph{feedback mechanism between
  the agents' behaviour and their in-group's structure}.

\bigskip 

In the context of local cultures, we are mainly interested in whether the
dynamics lead to consensus or in quantifying the degree of
heterogeneity in agent behaviour. We therefore
investigate how the results for the baseline model are affected by the
two extensions proposed here, namely the evolving agent network
and its feedback on agent behaviour. As known, equilibrium outcomes depend on
the key parameter $\varepsilon$ which distinguishes between
open-mindedness and narrow-mindedness of agents. Similar to the baseline
model, high values of $\varepsilon$ favour consensus or a small number of
unrelated population subgroups (=components).\footnote{For both models, this
does not hold in
  general as the probability of consensus and the average maximum
  component size as a function of $\varepsilon$ are not monotonically
  increasing (see \citet{ben-naim03} for the baseline model).} Note that in
equilibrium, the dynamics always partition the agent population into a
certain number of network \emph{components}, where all agents within a
component share the same
behaviour. Obviously, the difference between the behaviour in any two
components is at least the threshold $\varepsilon$ as interaction between agents
from different components would still be possible otherwise. Further, each
agent's in-group coincides with her respective network component: Agents within
a component are fully connected but have no links to outside agents.

\section{Findings}
\label{sec:results}

As explained before, agents need to agree on a desirable business behaviour to
allow for a local culture and cluster benefits to emerge. This consensus
results from past interactions and the successful behaviour therein.  
The strength of any local culture thus relates to the spread of a
specific behaviour in the population.  Therefore one quantity measured in our
simulations is the \emph{frequency of consensus} among all
agents, i.e. how often all agents exhibit the same
behaviour in equilibrium. As the behaviour profile does not converge within a
finite number of timesteps, we call a behaviour profile $x$ \textit{consensus
profile} if the maximum distance between two agents' respective behaviours is at
most $\varepsilon$.  In this case, all agents' behaviour will finally 
converge to the mean behaviour in the population. Similarly, we can
define a sufficient condition for non-convergence. First, there must be two
components whose distance is at least $\varepsilon$, i.e. if there exist agents
$i$ and $j$ with $|x_i-x_j|\geq\varepsilon$ and there is no other agent whose
behaviour is between $x_i$ and $x_j$. Second, we require that there are no
links between these $\varepsilon$-seperated components. In this case, the
respective limit behaviours for $i$ and $j$ even for large $t$ cannot coincide.
However, we should also take into account to what extent there
is a shared behaviour among the agents in case of no consensus. We
measure this by the \emph{size of the largest component} of agents
with identical behaviour. For example, a situation where 95\
agents share the same behaviour is much closer to consensus than one
where the population of agents splits into three equally sized components
with different behaviour.

We analyse the local cultures model with respect to these quantities by
means of computer simulations and compare the results to the baseline
model that has no feedback between behaviour and network. The key
parameter varied is $\epsilon$, high values of which characterize the
\emph{openmindedness} of the agents.  
\begin{figure}[htbp]
  \centering
  \subfigure[$n=50$, $\mu=0.5$]{
  \includegraphics[width=0.45\textwidth]{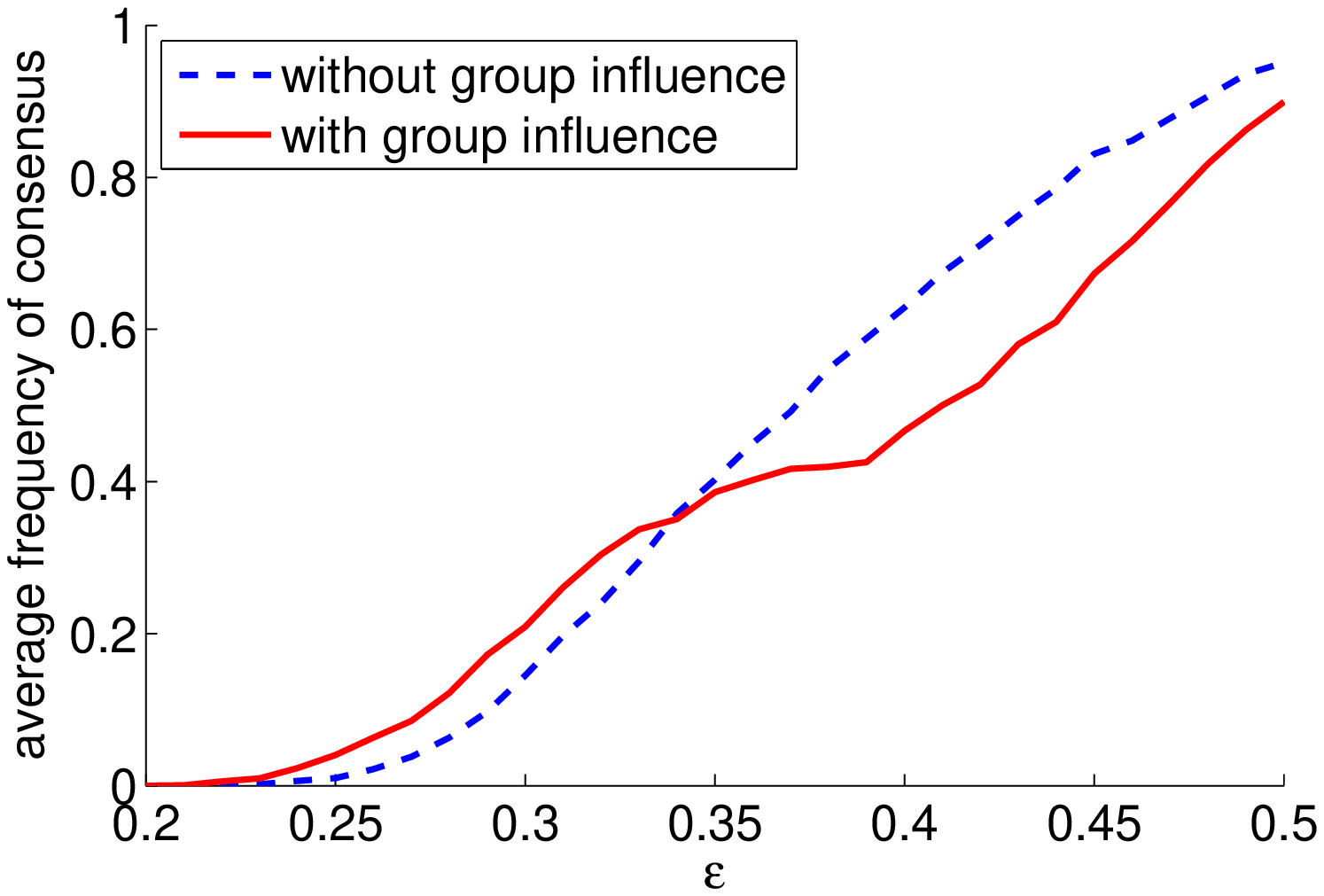}
  \label{subfig:consensus_freq_n_50_mu_0.5} } \subfigure[$n=50$, $\mu=0.1$]{
  \includegraphics[width=0.45\textwidth]{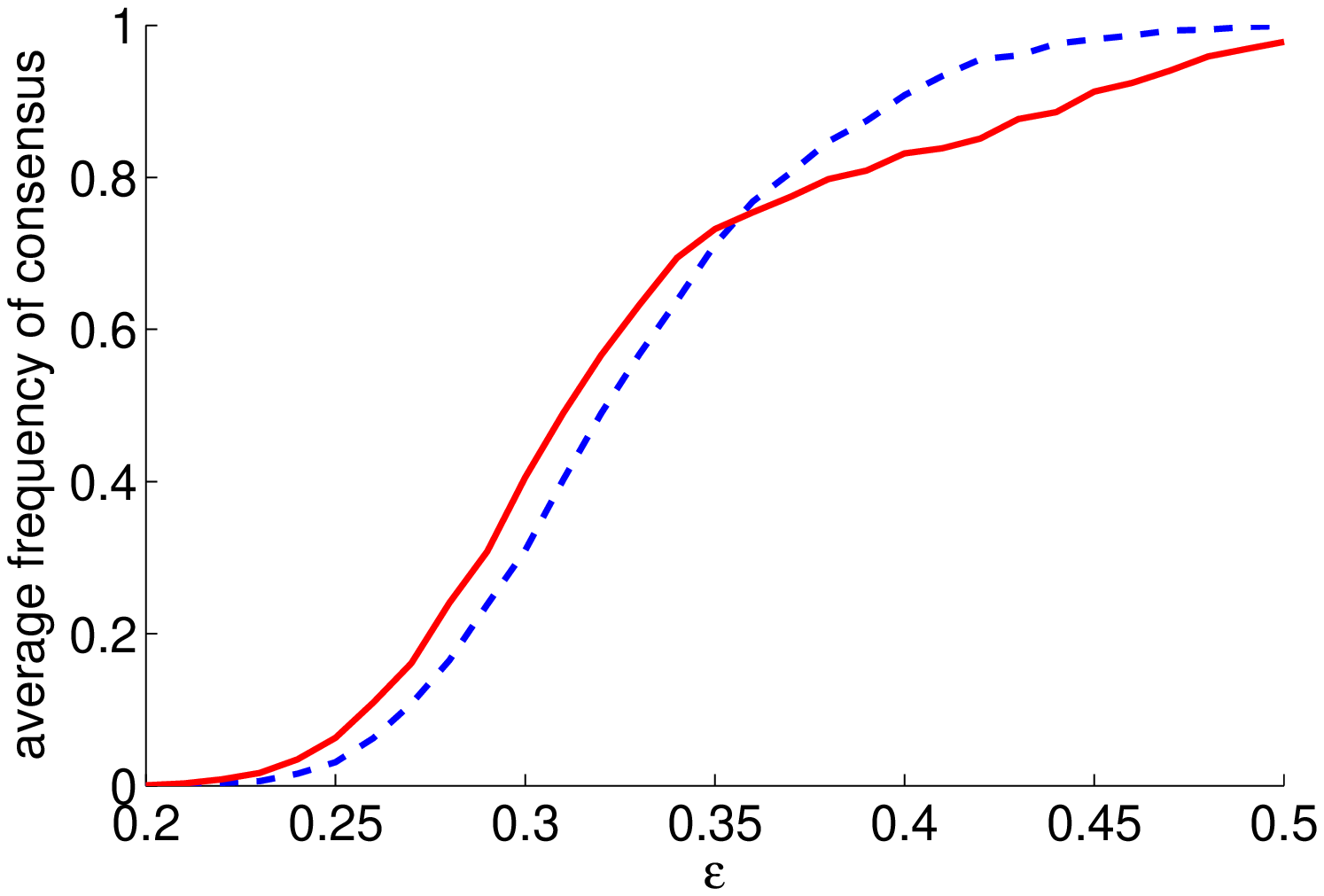}
  \label{subfig:consensus_freq_n_50_mu_0.1} }
  \subfigure[$n=100$, $\mu=0.5$]{
\includegraphics[width=0.45\textwidth]{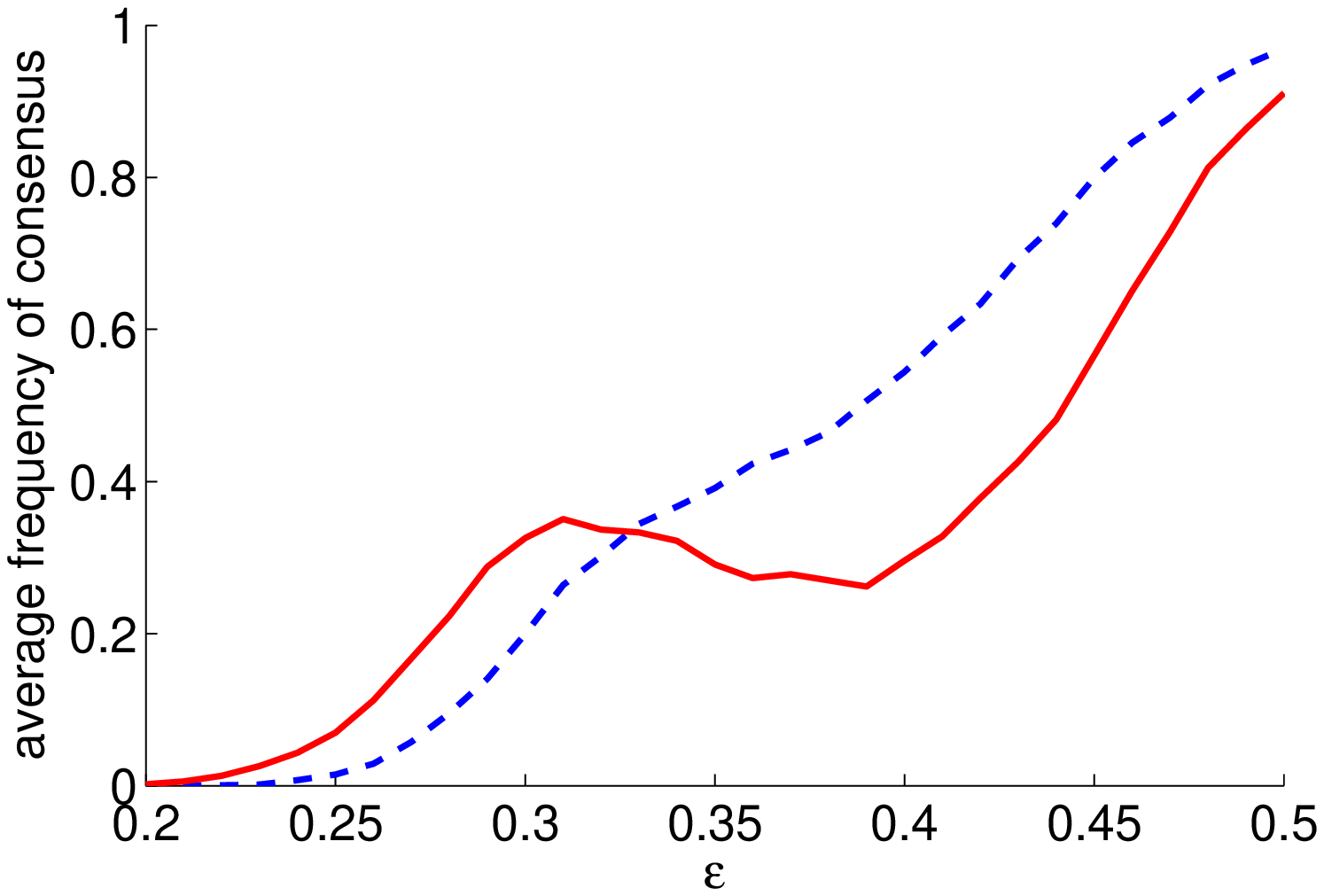}
    \label{subfig:consensus_freq_n_100_mu_0.5} } \subfigure[$n=100$, $\mu=0.1$]{
\includegraphics[width=0.45\textwidth]{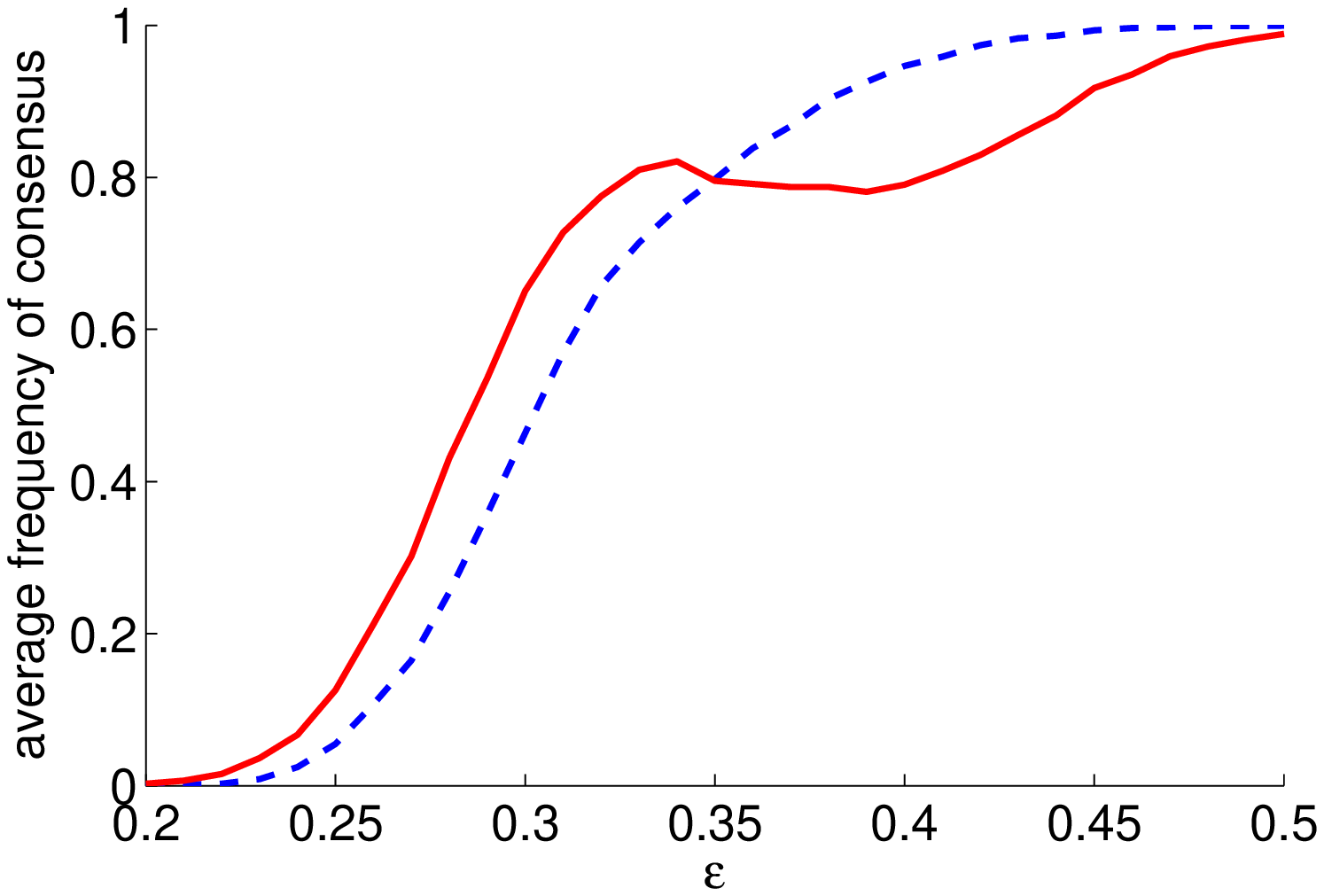}
    \label{subfig:consensus_freq_n_100_mu_0.1} }
    \subfigure[$n=500$, $\mu=0.5$]{
   
\includegraphics[width=0.45\textwidth]{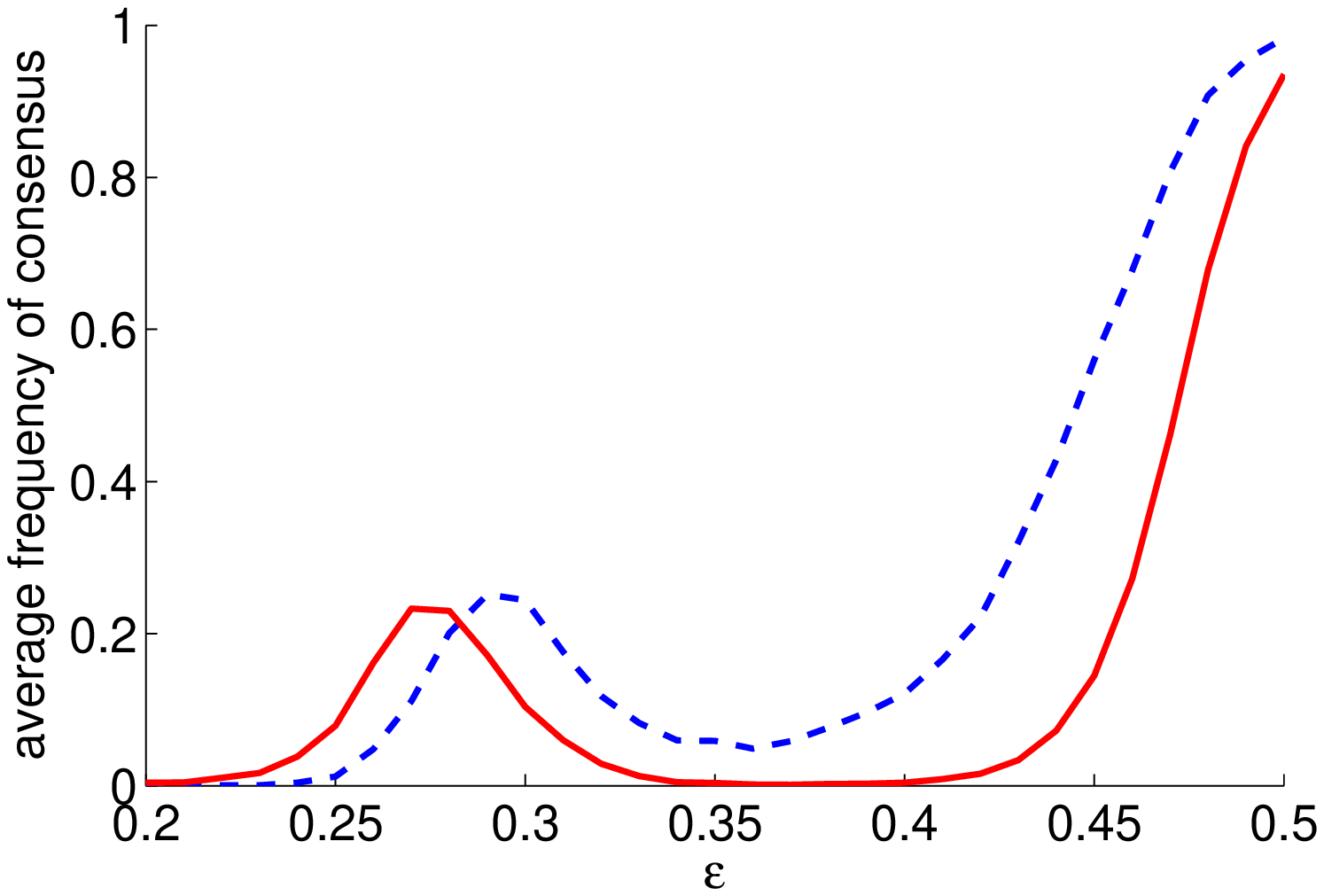}
    \label{subfig:consensus_freq_n_500_mu_0.5} } \subfigure[$n=500$, $\mu=0.1$]{
   
\includegraphics[width=0.45\textwidth]{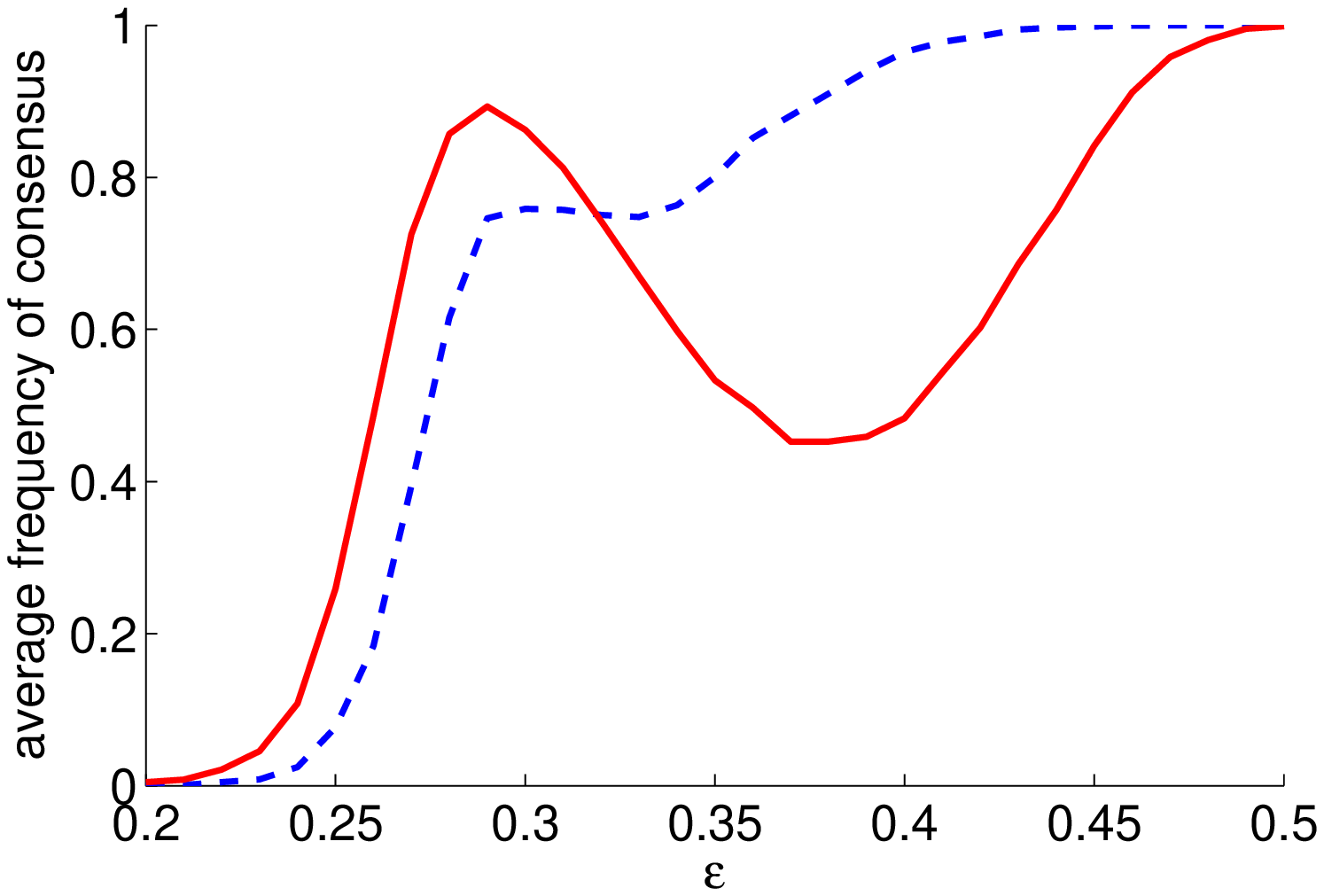}
    \label{subfig:consensus_freq_n_500_mu_0.1} }
  \caption{Average frequency of consensus dependent on $\epsilon$ for 5000 runs
    and different values of $n$ and $\mu$. The agents' initial behaviour is
    random according to a uniform distribution, the initial network is empty.
For narrow-minded agents (small $\varepsilon$), the group influence fosters
consensus while for open-minded agents (high $\varepsilon$), the probability of
indentical behaviour is lower than in the baseline model without group
influence.
  \label{fig:consensus_freq}}
\end{figure}

\begin{figure}[htbp]
  \centering
    \subfigure[$n=50$, $\mu=0.5$]{
    \includegraphics[width=0.45\textwidth]{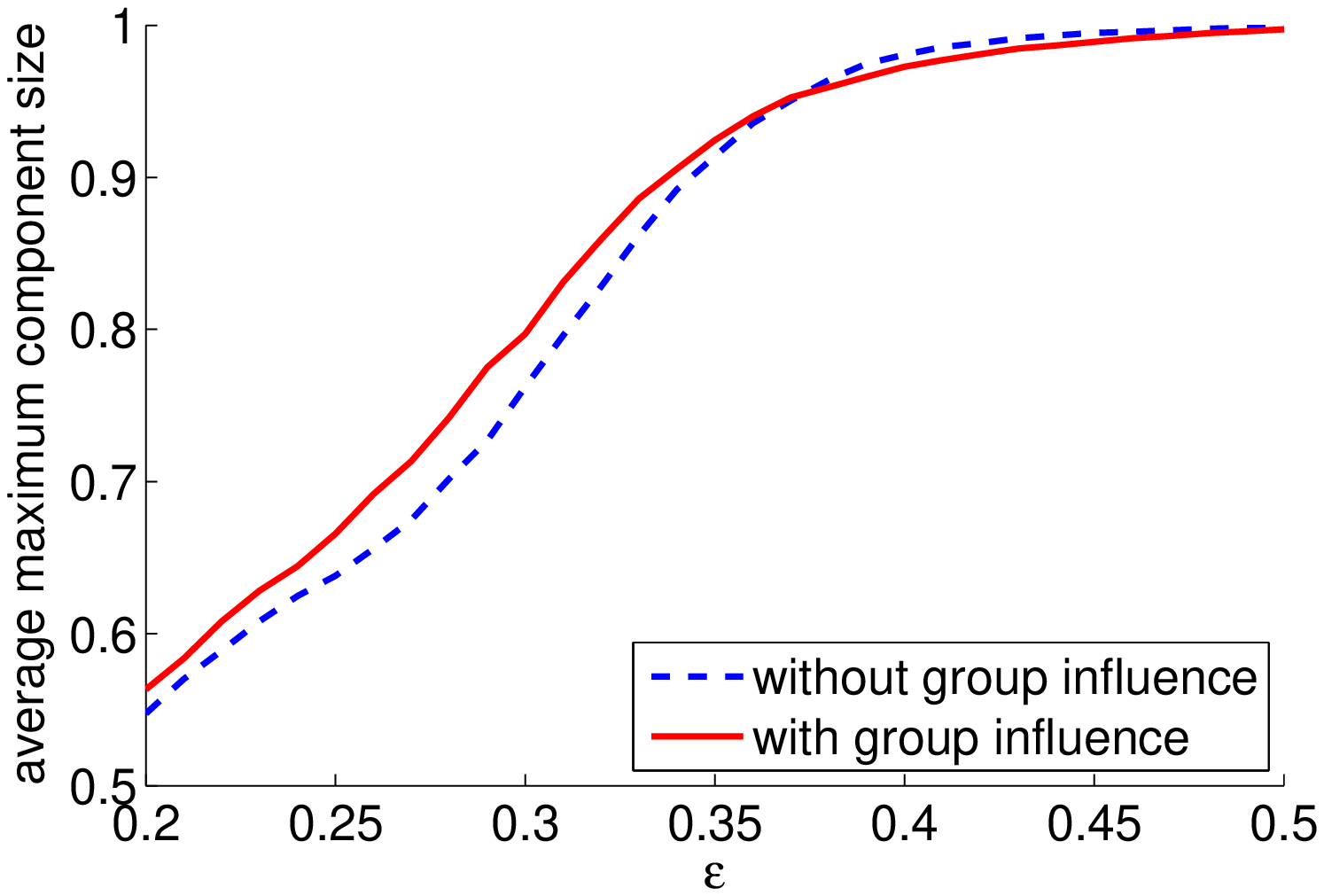}
    \label{subfig:max_cluster_n_50_mu_0.5} } \subfigure[$n=50$, $\mu=0.1$]{
    \includegraphics[width=0.45\textwidth]{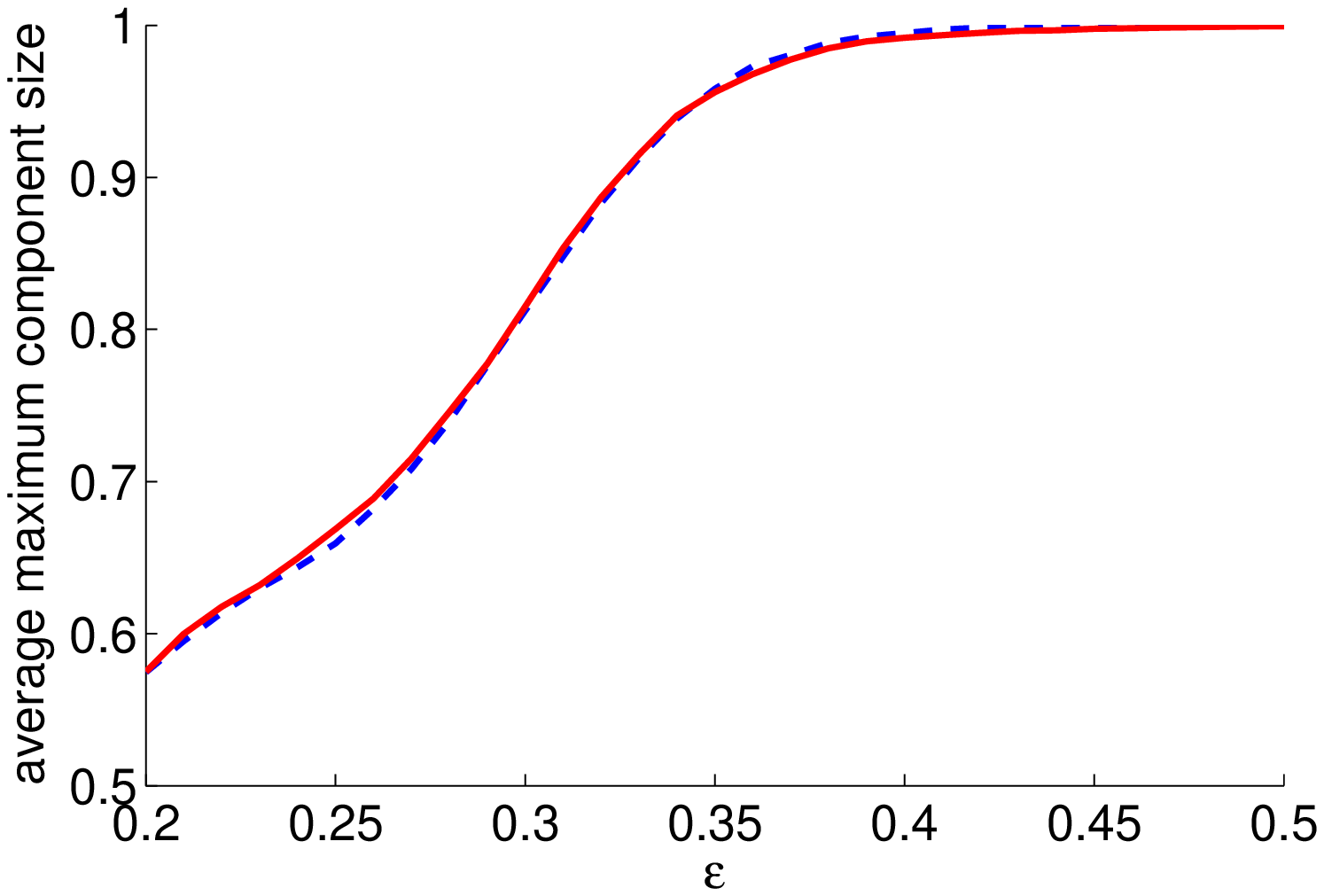}
    \label{subfig:max_cluster_n_50_mu_0.1} }
    \subfigure[$n=100$, $\mu=0.5$]{
    \includegraphics[width=0.45\textwidth]{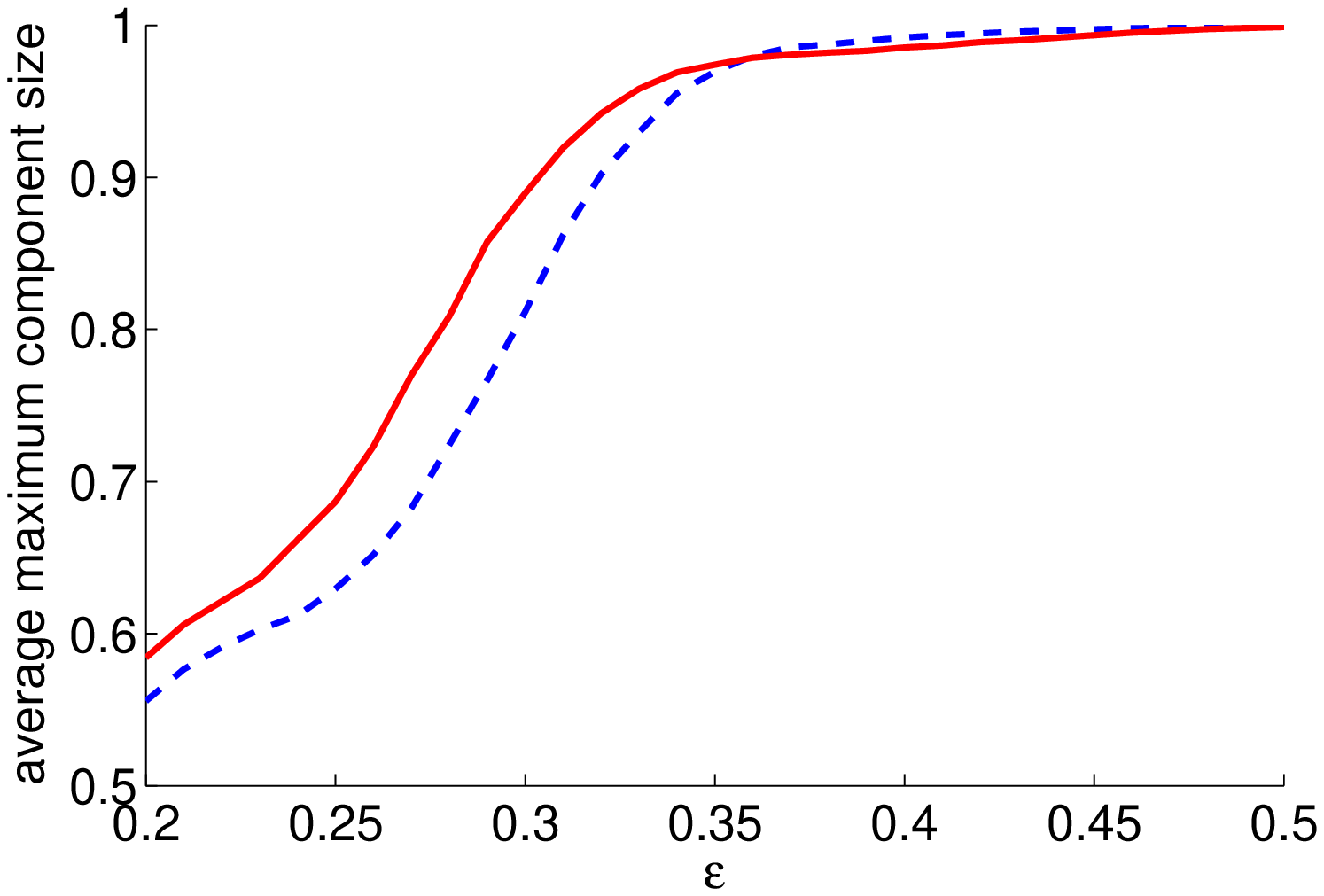}
    \label{subfig:max_cluster_n_100_mu_0.5} } \subfigure[$n=100$, $\mu=0.1$]{
    \includegraphics[width=0.45\textwidth]{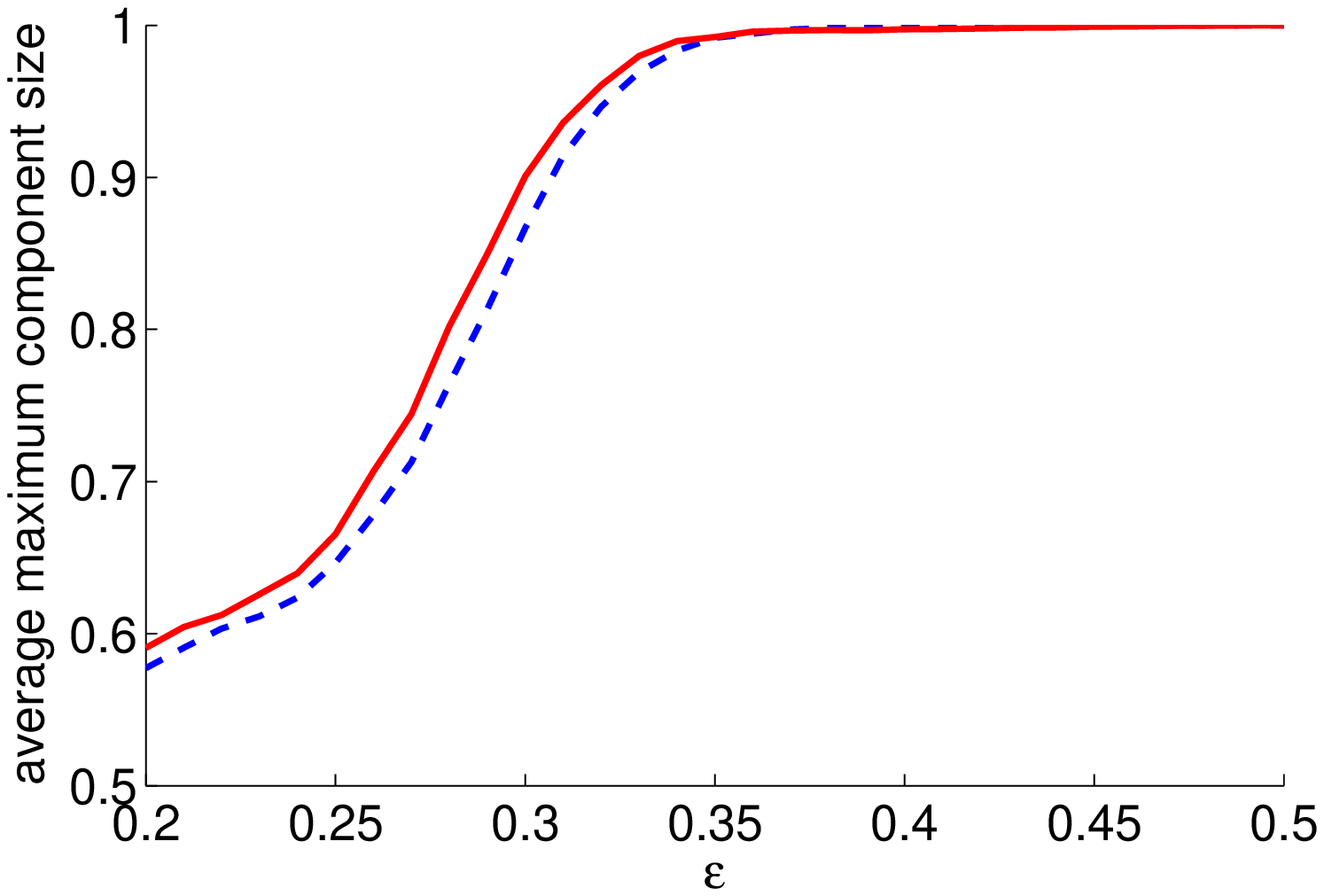}
    \label{subfig:max_cluster_n_100_mu_0.1} }
    \subfigure[$n=500$, $\mu=0.5$]{
    \includegraphics[width=0.45\textwidth]{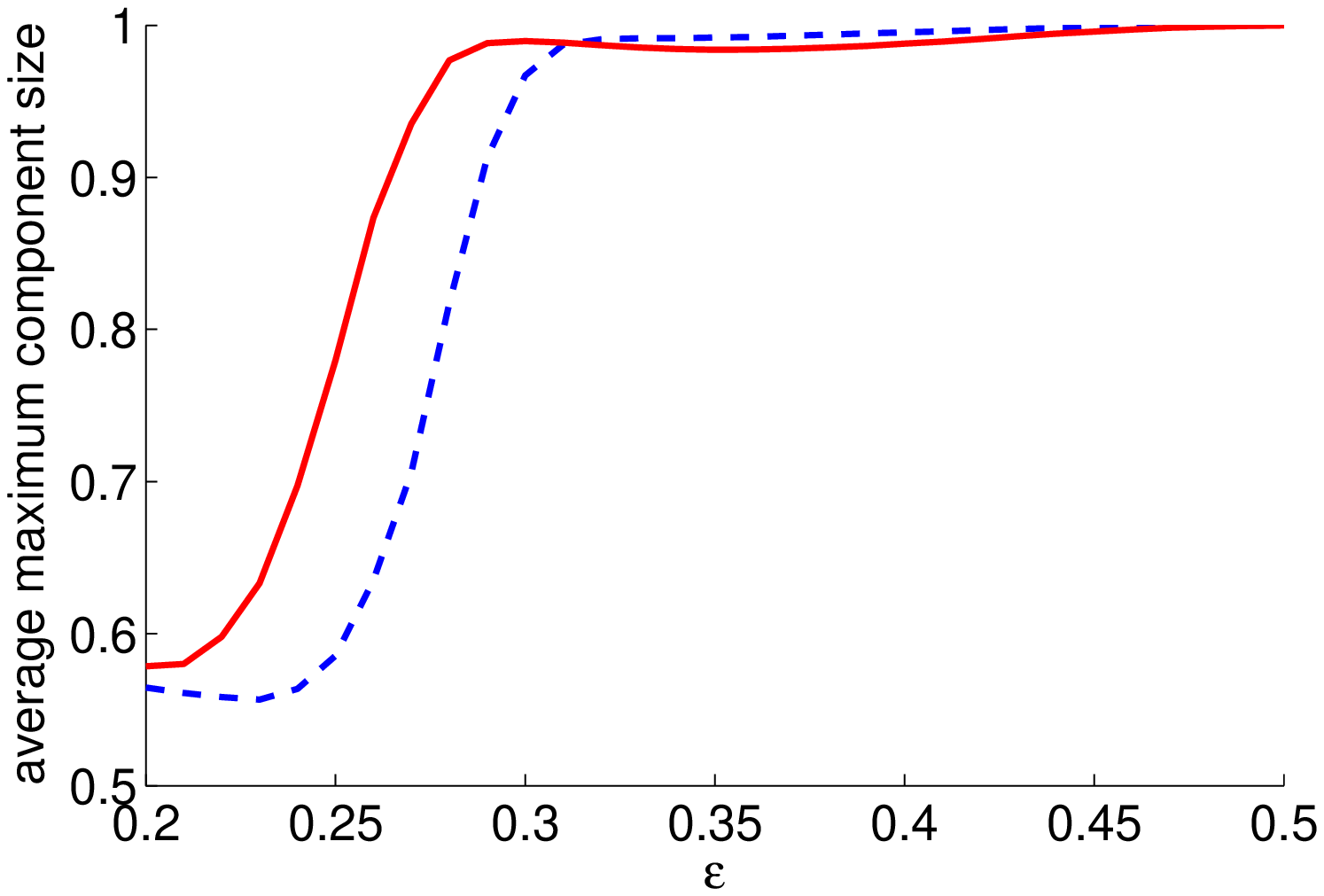}
    \label{subfig:max_cluster_n_500_mu_0.5} } \subfigure[$n=500$, $\mu=0.1$]{
    \includegraphics[width=0.45\textwidth]{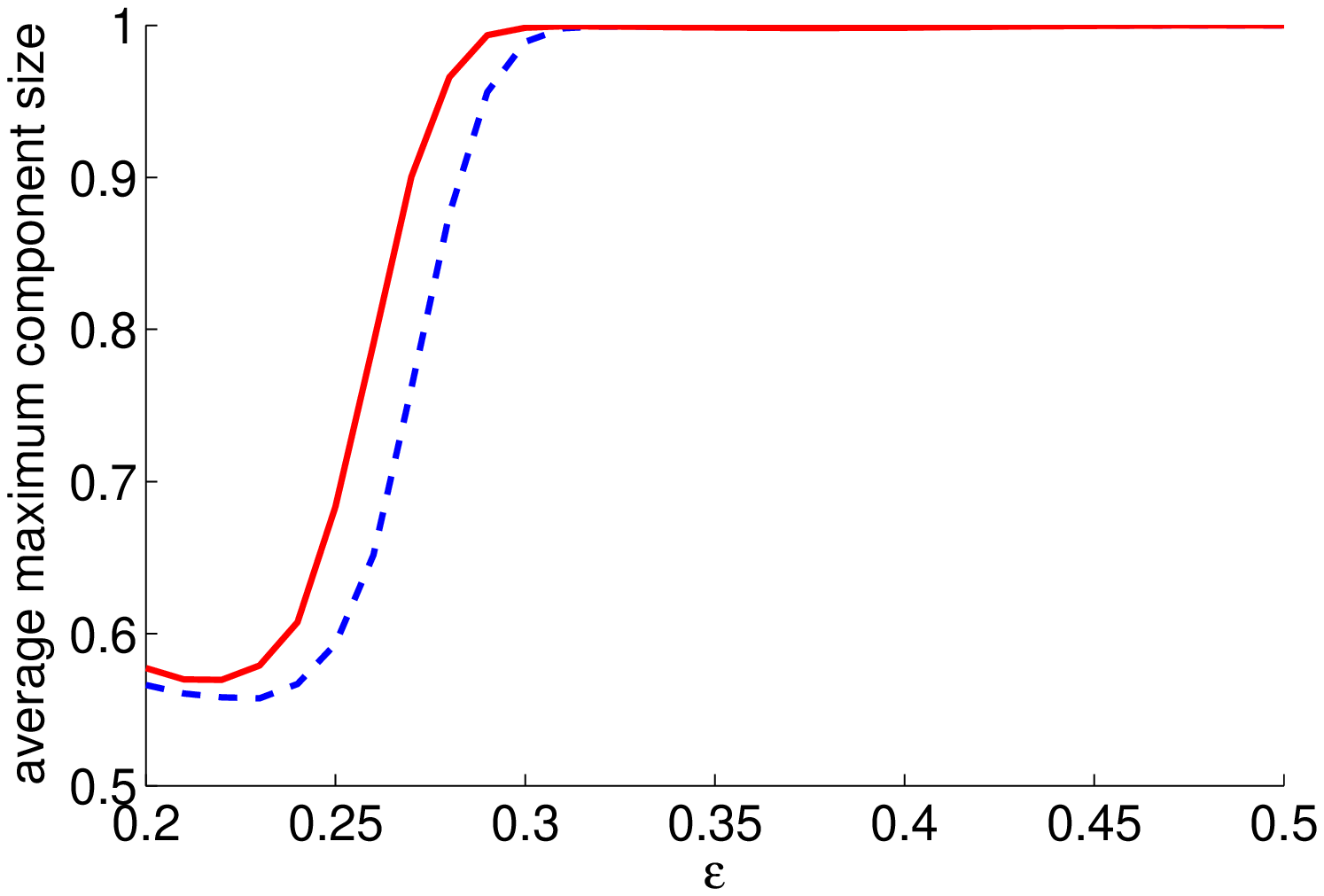}
    \label{subfig:max_cluster_n_500_mu_0.1} }
  \caption{Average maximum relative component
    size dependent on $\epsilon$ for 5000 runs and
    different values of $n$ and $\mu$. The agents' initial behaviour is random
    according to a uniform distribution, the initial network is empty. For
narrow-minded agents (small $\varepsilon$), our model increases the
average maximum component size compared to the baseline model without group
influence.
For open-minded agents (high $\varepsilon$), there is almost no difference
between the two models with respect to the average maximum component size.
  \label{fig:max_cluster}}
\end{figure}

\begin{figure}[htbp]
  \centering
    \subfigure[$n=50$, $\mu=0.5$]{
   
\includegraphics[width=0.45\textwidth]
{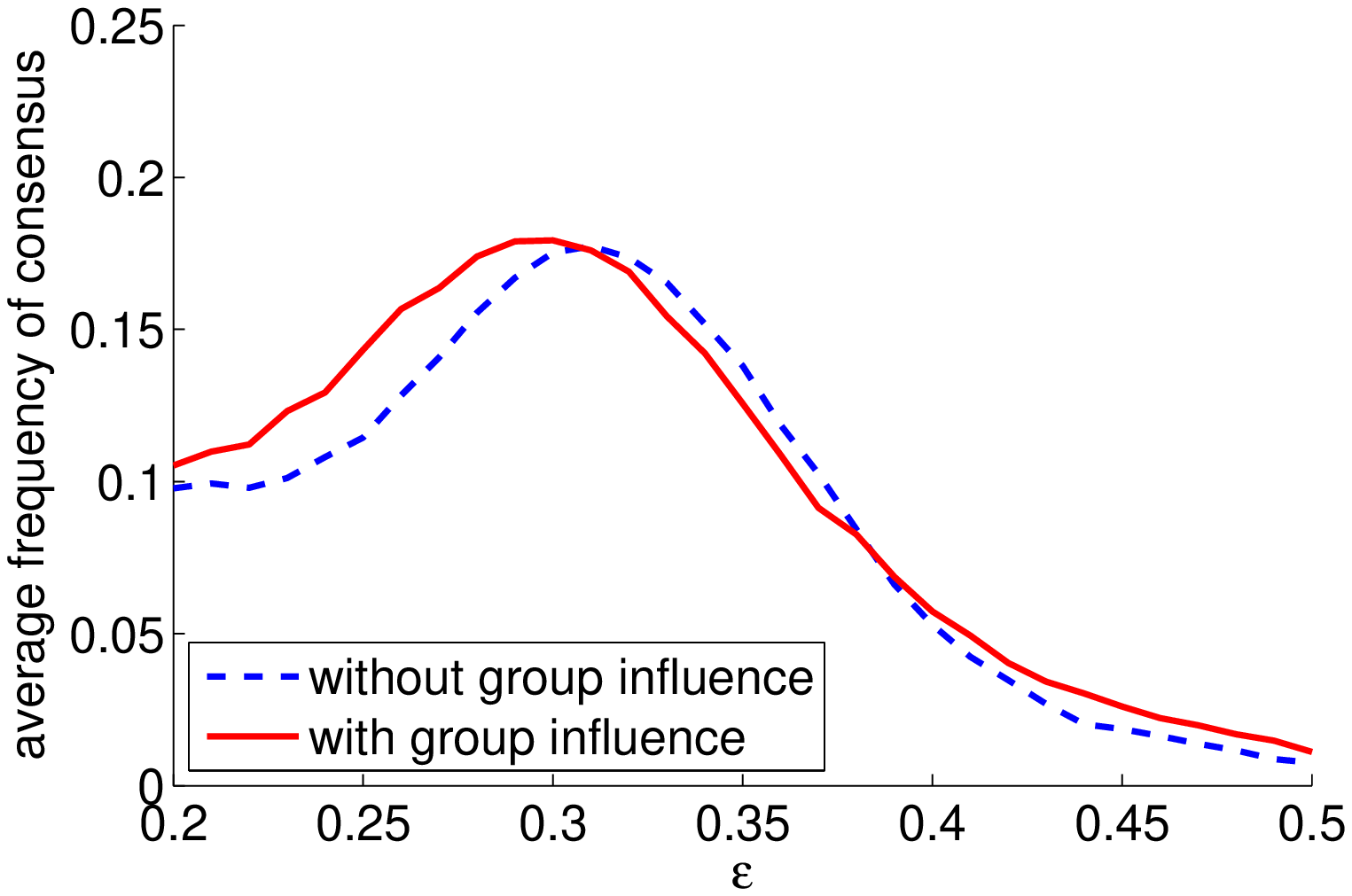}
    \label{subfig:max_cluster_stdev_n_50_mu_0.5} } \subfigure[$n=50$,
$\mu=0.1$]{
   
\includegraphics[width=0.45\textwidth]
{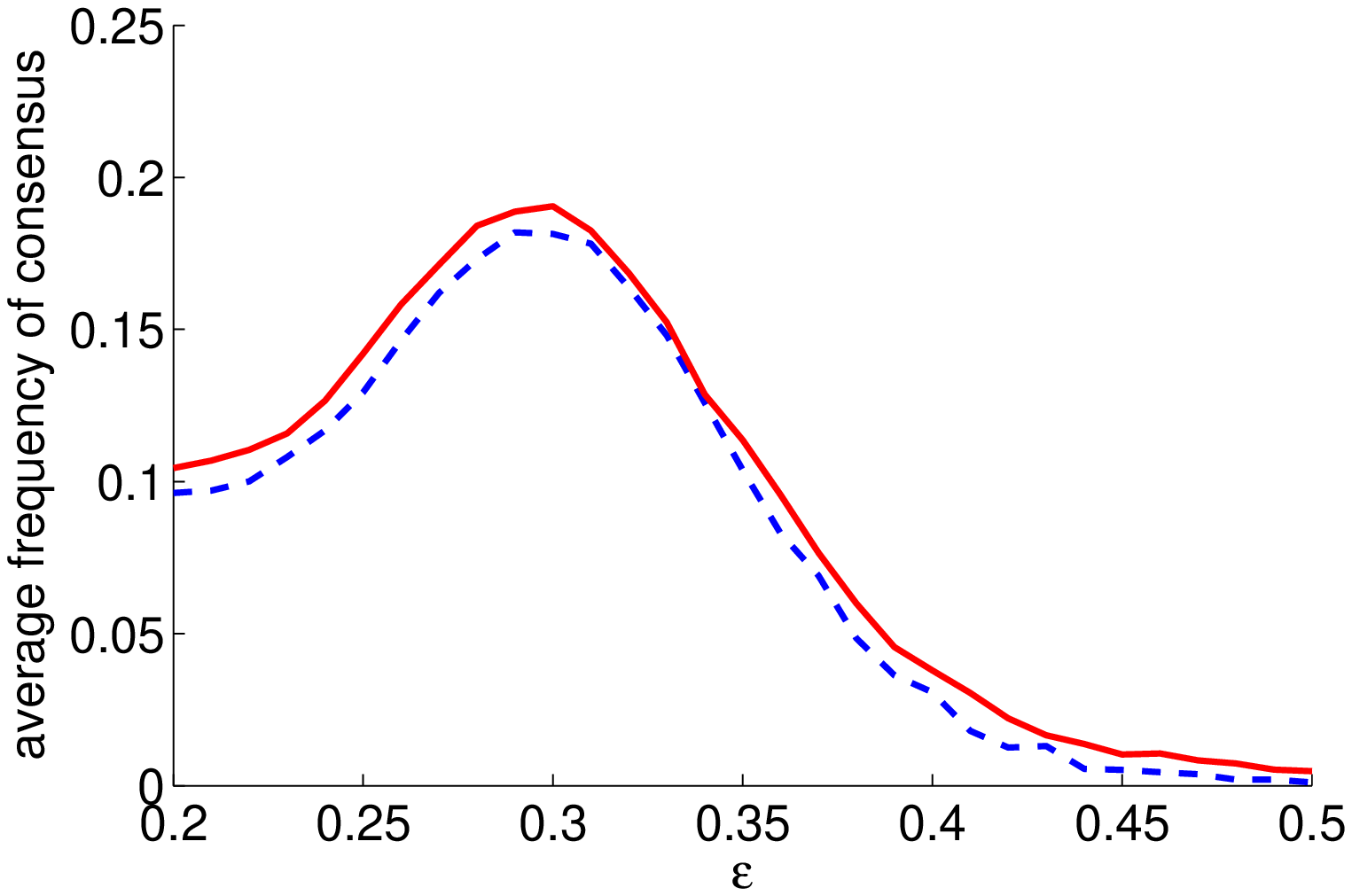}
    \label{subfig:max_cluster_stdev_n_50_mu_0.1} }
    \subfigure[$n=100$, $\mu=0.5$]{
   
\includegraphics[width=0.45\textwidth]
{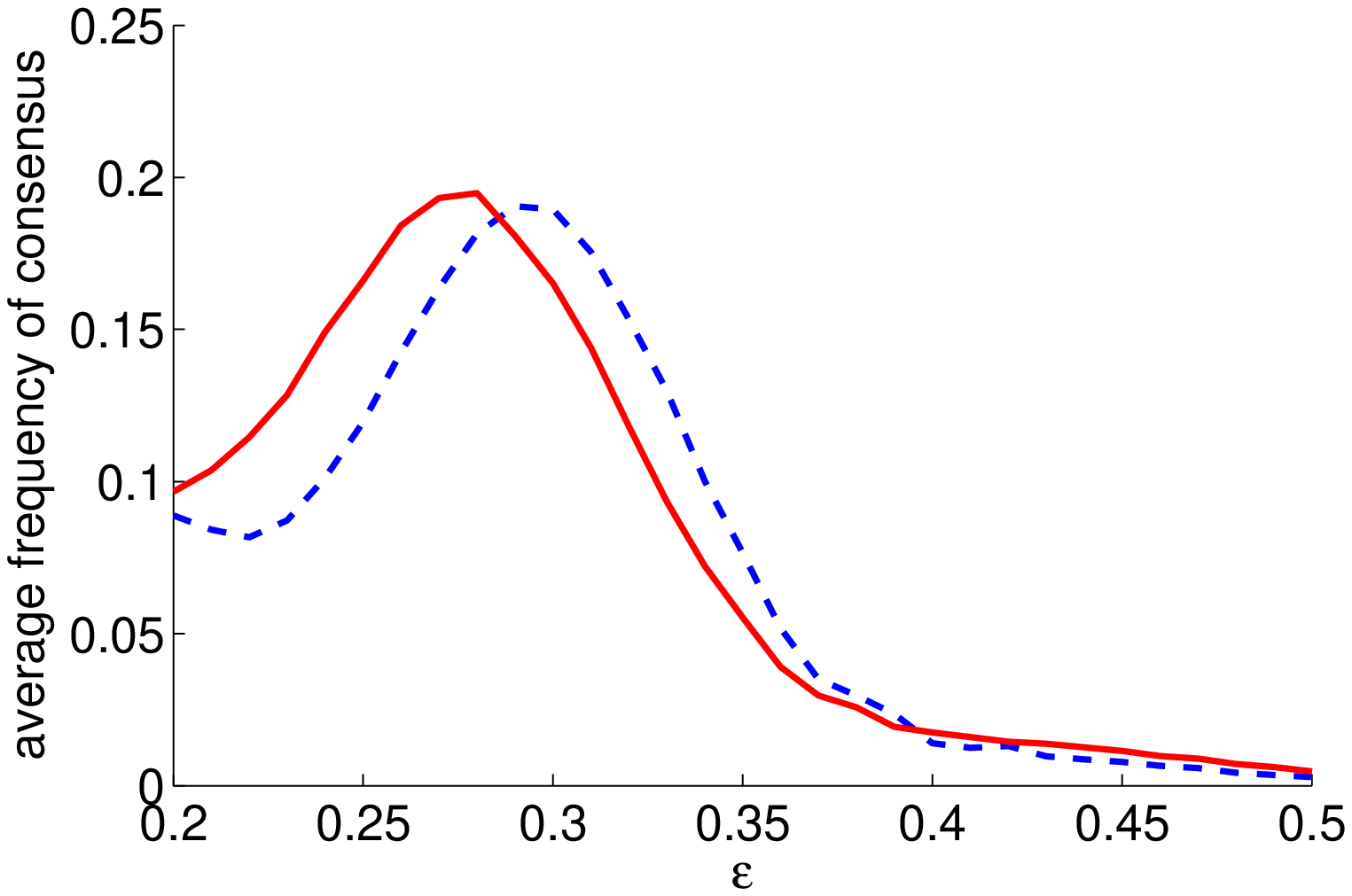}
    \label{subfig:max_cluster_stdev_n_100_mu_0.5} } \subfigure[$n=100$,
$\mu=0.1$]{
   
\includegraphics[width=0.45\textwidth]
{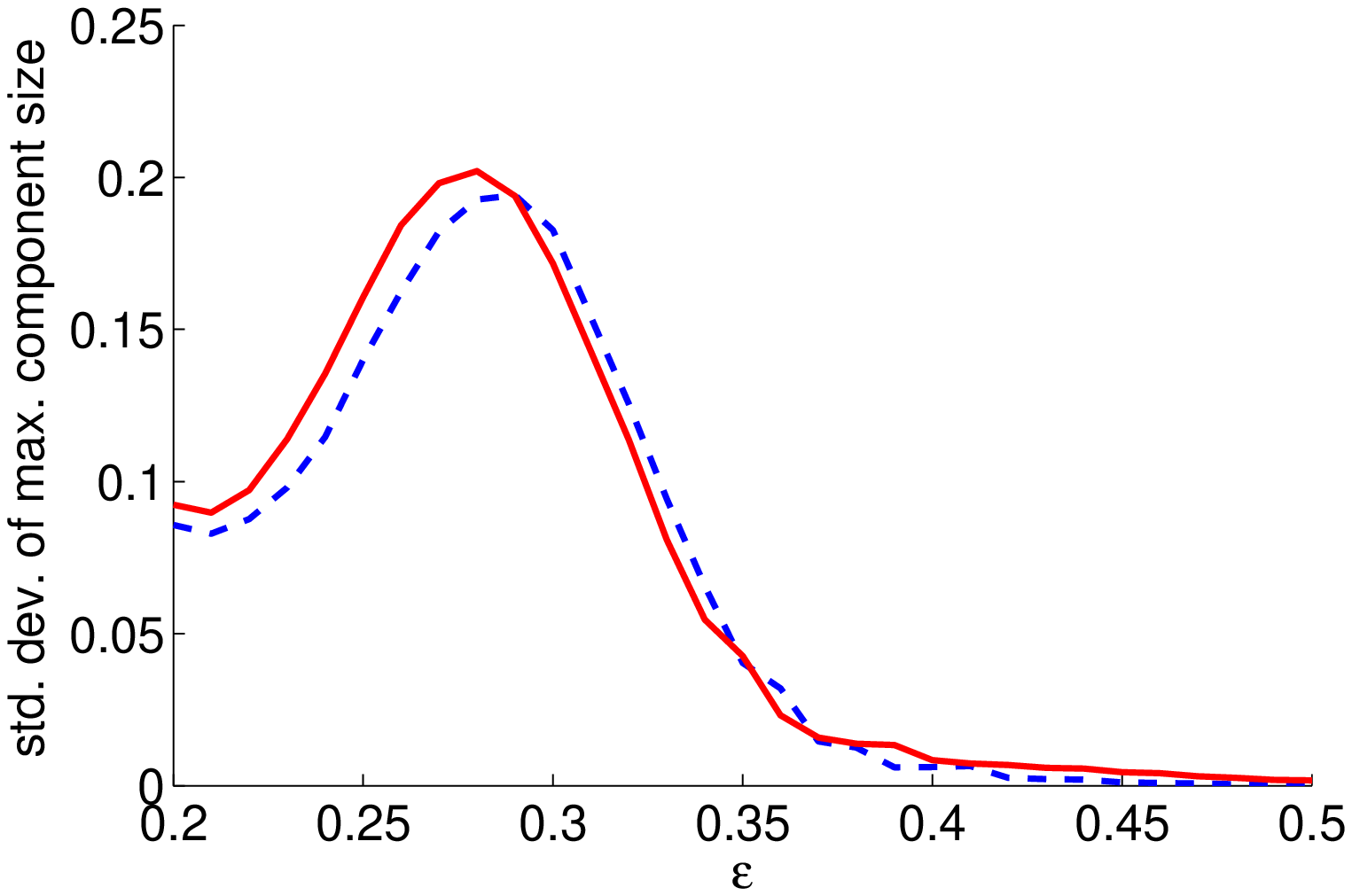}
    \label{subfig:max_cluster_stdev_n_100_mu_0.1} }
    \subfigure[$n=500$, $\mu=0.5$]{
   
\includegraphics[width=0.45\textwidth]
{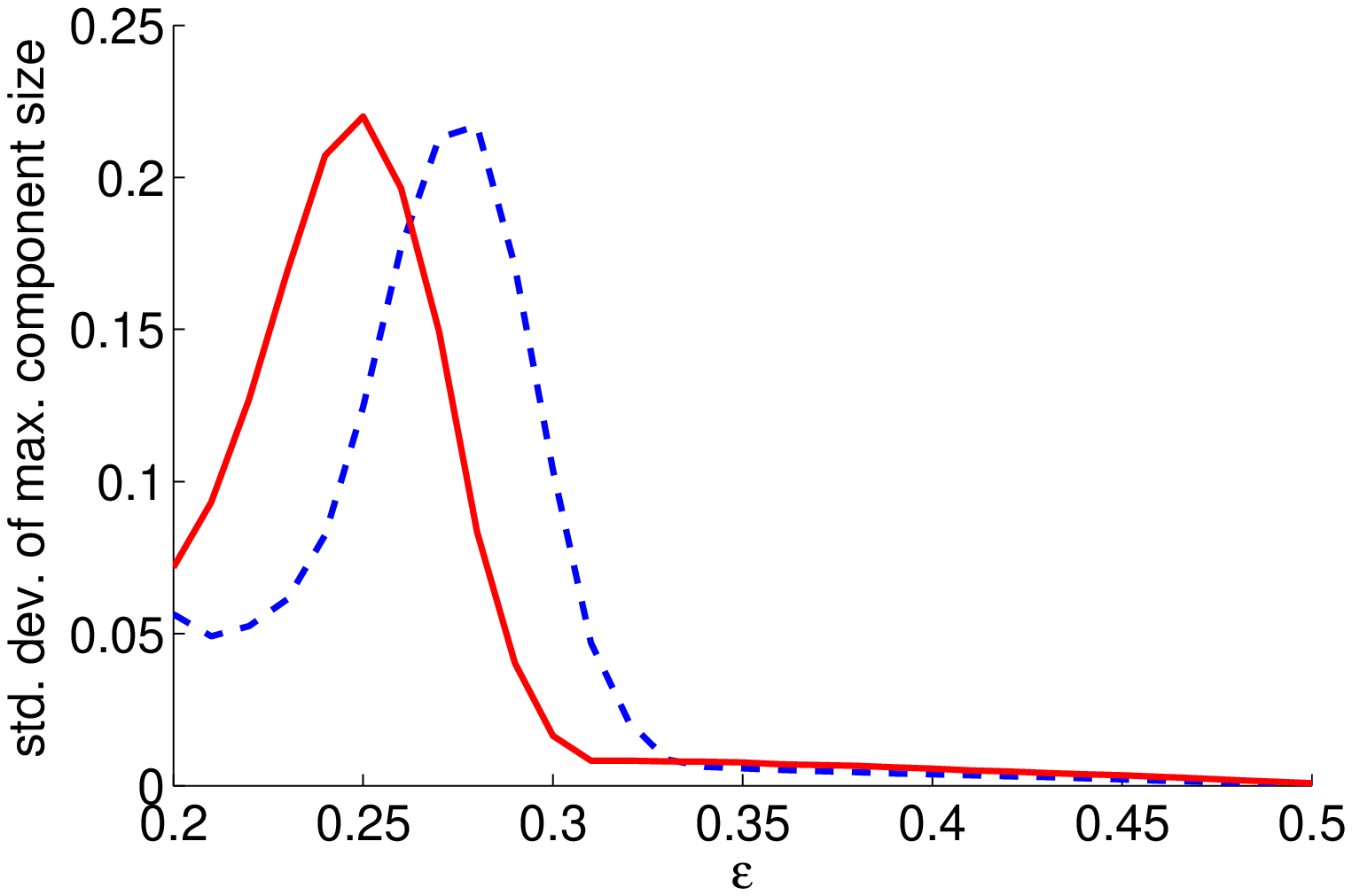}
    \label{subfig:max_cluster_stdev_n_500_mu_0.5} } \subfigure[$n=500$,
$\mu=0.1$]{
   
\includegraphics[width=0.45\textwidth]
{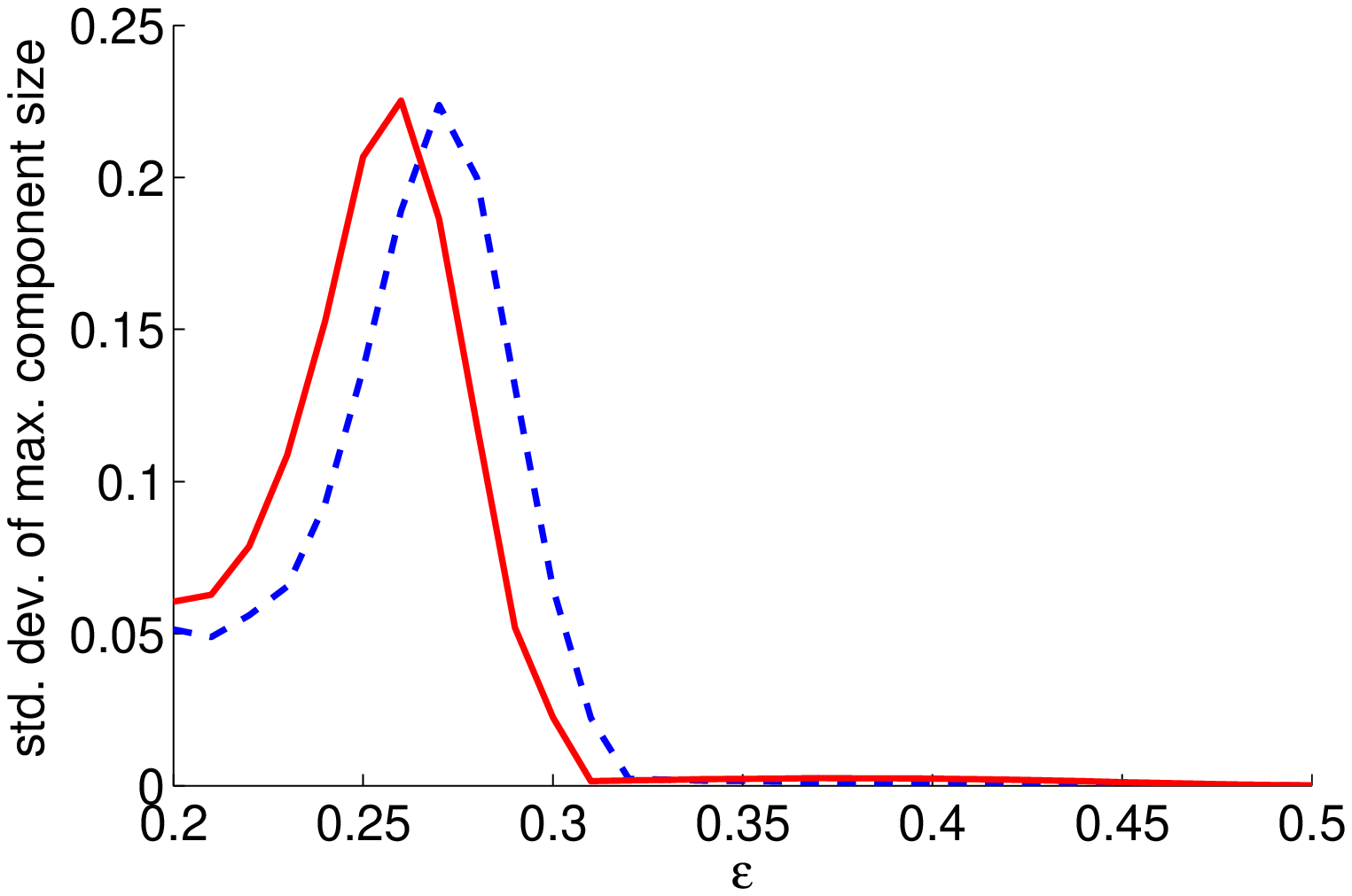}
    \label{subfig:max_cluster_stdev_n_500_mu_0.1} }
  \caption{Standard deviation of the maximum relative group
    size dependent on $\epsilon$ for 5000 runs and
    different values of $n$ and $\mu$. The agents' initial behaviour is random
    according to a uniform distribution, the initial network is empty.
  \label{fig:max_stdev_cluster}}
\end{figure}

The main result relates to the question whether the feedback mechanism
between the agents' behaviour and their network structure (group influence) is
beneficial in the sense that a common behaviour is fostered. Our
simulations (see Figure \ref{fig:consensus_freq} and
Figure \ref{fig:max_cluster})
show that for more narrow-minded agents, i.e. small
thresholds $\varepsilon$, the group influence
results in both a higher average frequency of consensus and a larger
maximum component size as compared to the baseline model. Hence, the
mechanism introduced in our model increases the likelihood of consensus
formation. 

This result, interestingly and counterintuitively, changes for larger
thresholds and therefore more open-minded agents. Here, the feedback mechanism
\emph{weakens} the emergence of a local culture in general. Agents subject
to group influence reach less consensus on average than in the baseline
model. We note, however, that the mechanism's effect differs between the
two measures: While the frequency of consensus is
significantly decreased
(Figure \ref{fig:consensus_freq}), the effect on the maximum component size
is much smaller (Figure \ref{fig:max_cluster}). 
To explain the influence of agents open-mindedness, we argue that the feedback
mechanism in the local cultures model
implies two opposed effects compared to the baseline model. On the one hand,
agents with ``extreme'' initial behaviour (i.e. an initial behaviour close to
zero or one) are less likely to interact with other agents and are therefore
more likely to stay in that border area.  The longer these agents remain in
isolation (i.e. without interacting), the denser the network of other
agents becomes, implying more averaging of behaviour in determining the
effective behaviour of these networked agents. As more averaging leads to
values closer to the mean, there are fewer and fewer agents within the
interaction range of any isolated agent (as compared to the baseline
model), and full consensus becomes more unlikely.

\begin{figure}[htbp]
  \centering
  \subfigure[$t=240$]{
    \includegraphics[width=0.33\textwidth]{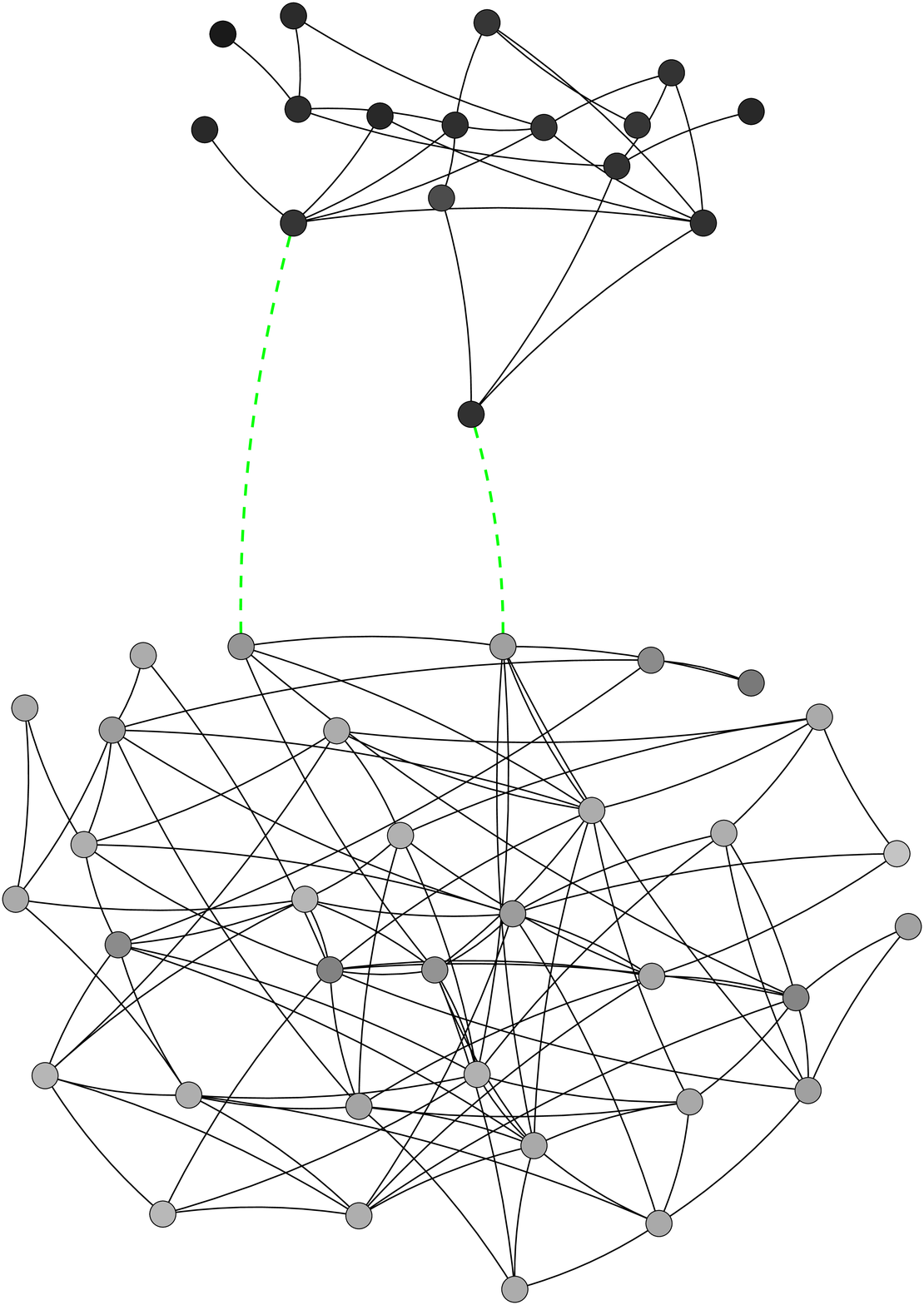}
    \label{subfig:coalescence1}
  }
  \subfigure[$t=530$]{
    \includegraphics[width=0.33\textwidth]{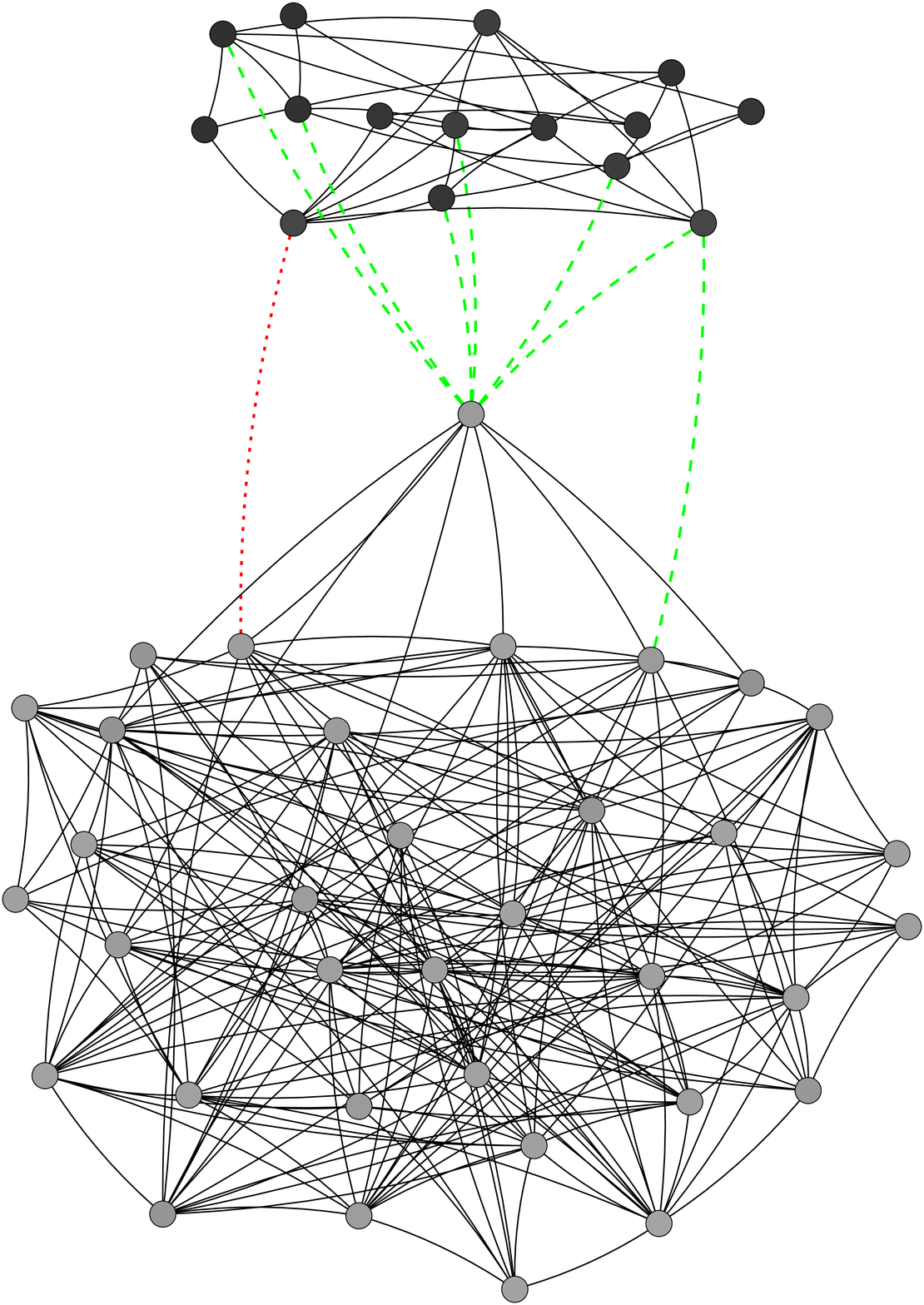}
    \label{subfig:coalescence2}
  }
  \subfigure[$t=750$]{
    \includegraphics[width=0.33\textwidth]{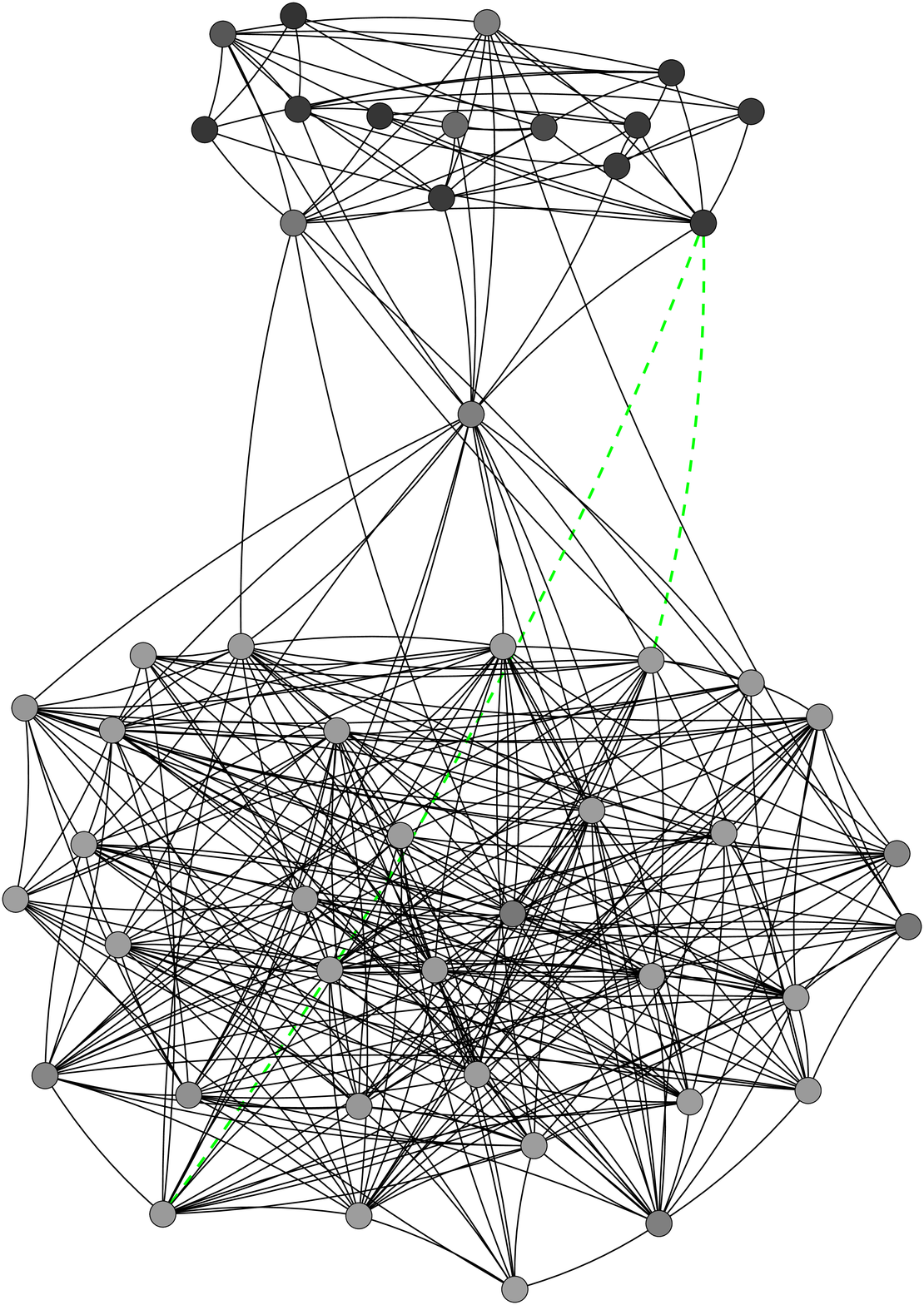}
    \label{subfig:coalescence3}
  }
  \subfigure[$t=810$]{
    \includegraphics[width=0.33\textwidth]{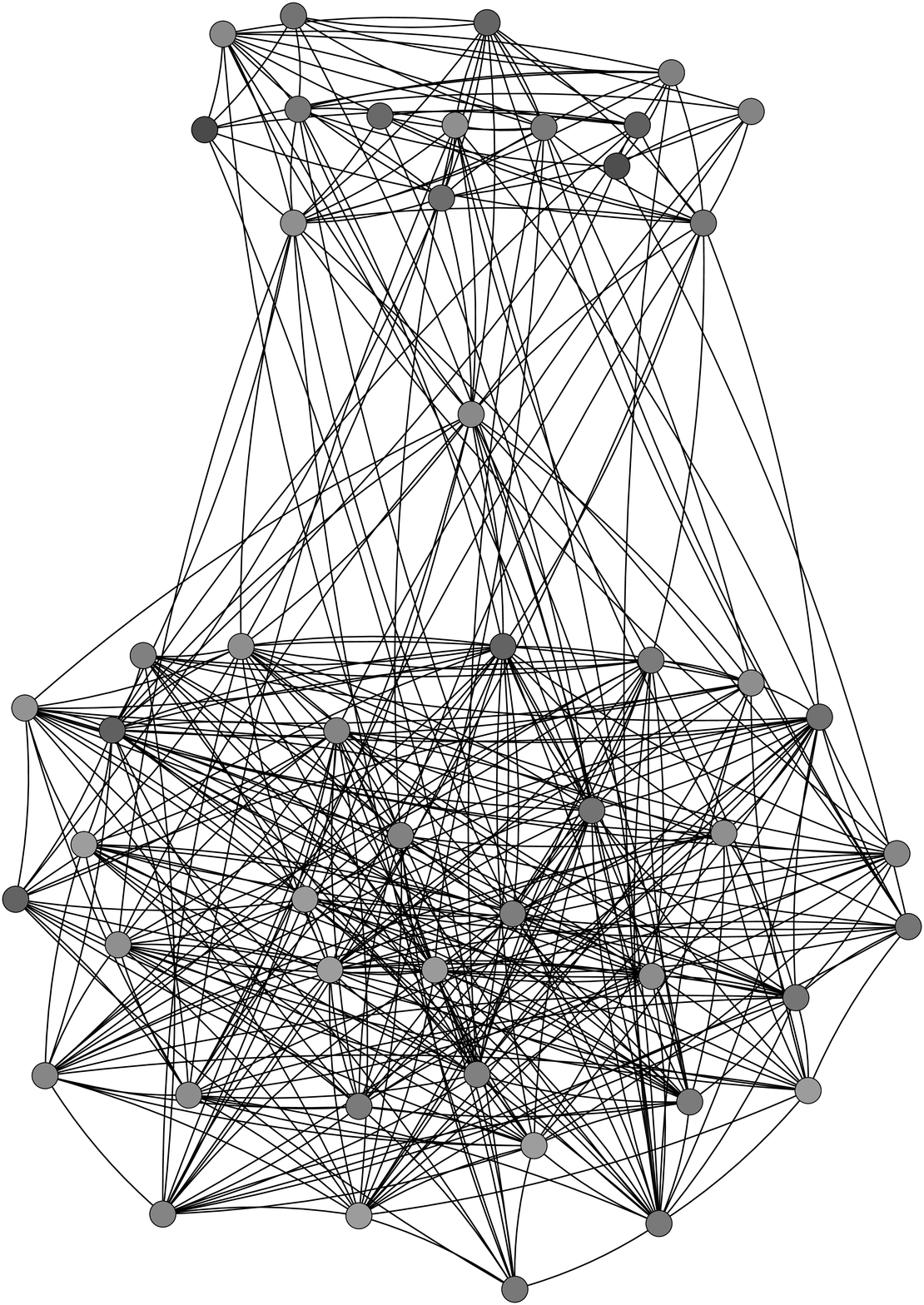}
    \label{subfig:coalescence4}
  }
  \caption{Network evolution for a simulation with 50 agents and
$\varepsilon=0.3$ at different timesteps. The agents' initial behaviour is
random according to a uniform distribution, the initial network is empty. A
node's colour indicates the respective agent's behaviour (white=0,
black=1). A green dashed link denotes that the respective agents' difference in
effective behaviour is below the threshold while their respective own
behaviours differ more than $\varepsilon$. Thus, such a link persists in
the local cultures model but would be deleted in the baseline model. A
red dotted link indicates that the respective agents' difference in effective
behaviour is above the threshold, i.e. the link would be deleted if the
respective agents were chosen at that timestep.}
  \label{fig:coales}
\end{figure}

On the other hand, the feedback mechanism fosters consensus by increasing
the coalescence of subpopulations with different behaviour, i.e.
components within the network. To illustrate this, consider the
simulation depicted in Figure \ref{fig:coales}.\footnote{A video of this
simulation can be found at \texttt{
http://web.sg.ethz.ch/publications/local\_cultures/\\
web-cultures.html}.
For a dynamic
network layout, we use the \emph{arf} algorithm \citep{geipel07}.}
In Figure \ref{subfig:coalescence1}, there are two nearly separated
components in our model, the upper, smaller one with a higher average behaviour,
the lower, larger one with a lower average behaviour. The two links that
connect these components would not persist in the baseline model as the
respective nodes' difference in terms of their own behaviour is above the
threshold. However, as this is not the case for their effective
behaviour, the involved agents can still interact in our model.  Hence,
the two agents from the upper component still influence agents in the
lower component by increasing their effective behaviour (compared to
their other neighbours whose behaviour is lower). For the same reason,
the two upper agents' effective behaviour is decreased by its neighbours
from the lower component.  Thus, they could establish further connections
to the lower component.  Nevertheless, interaction with agents from the
upper component would increase their behaviour and hence increase the
distance to their neighbours from the lower component.\footnote{This increase
  would have two reasons: first the increase of their own behaviour,
  second the decrease of their lower component's neighbours' influence on
  their effective behaviour as the share of lower component agents of the
  neighbourhood also decreases.} Therefore, whether the two components
stay connected and finally evolve to a complete graph or become separated
depends on which nodes are chosen in the near future, i.e. is a path
dependent process. Any interaction between the \emph{different}
components increases the probability of their coalescence, any
interaction within the \emph{same} component decreases that
probability. In our example, one agent can establish further connections
to the lower component (Figure \ref{subfig:coalescence2}) and in return
enables its neighbors from different components to interact (Figure
\ref{subfig:coalescence3}). Very quickly, more and more agents from the
different components interact, become more similar and finally make the
components coalesce (Figure \ref{subfig:coalescence4}). This effect of
coalescing components is also apparent in a higher variance of the 
maximum component size for narrow-minded agents as compared to the baseline
model (cf. Figure \ref{fig:max_stdev_cluster}).

Which of these effects decides over the strength of a local culture
depends on the threshold $\varepsilon$: For narrow-minded agents, the baseline
 model is generally more likely to  obtain several components instead of
consensus. Thus, in the local cultures model, the increased
probability of the coalescence of components increases both the frequency
of consensus and the maximum component size. For open-minded agents, this
effect vanishes because of the greater ex-ante likelihood of consensus in
the baseline model. In this situation, the effect of isolation of agents
with extreme behaviour comes into play: While both models favour
consensus in general, it is more likely for the local cultures model to find
agents at the spectrum's borders being separated from the other agents
because of the faster dynamics towards the center. Therefore, the
frequency of consensus is lower in this model. On the other hand there
are only few agents separated from the majority, so the maximum
component size is only slightly decreased by the feedback mechanism in
the local cultures model. Hence, if we only consider this quantity to
measure a local culture's magnitude, the feedback mechanism significantly
strengthens a local culture for narrow-minded agents and only slightly
weakens it for open-minded agents.

\begin{figure}[htbp]
  \centering
  \subfigure[without group influence]{
    \includegraphics[width=0.47\textwidth]{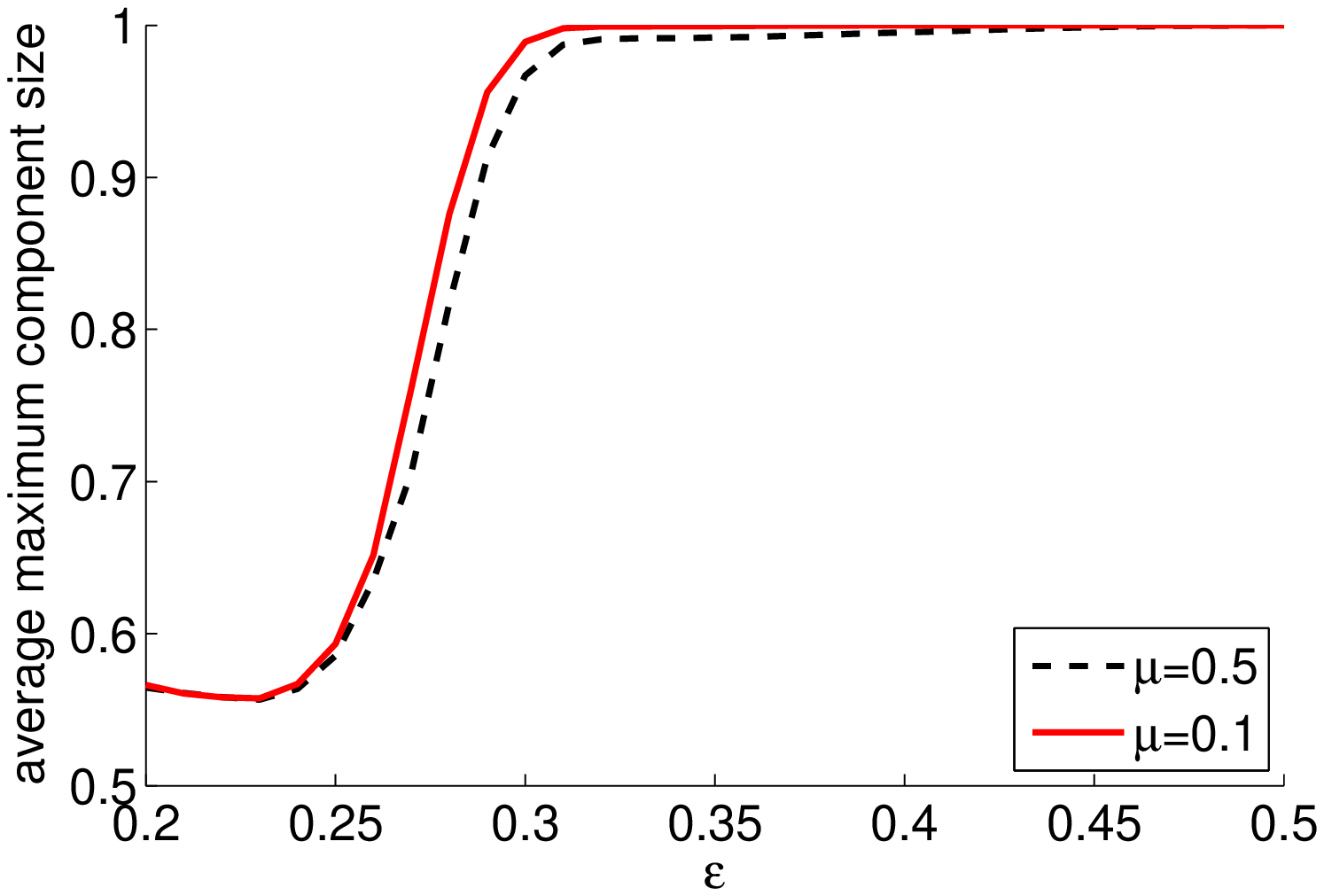}
    \label{subfig:compstat1}
  }
  \subfigure[with group influence]{
    \includegraphics[width=0.47\textwidth]{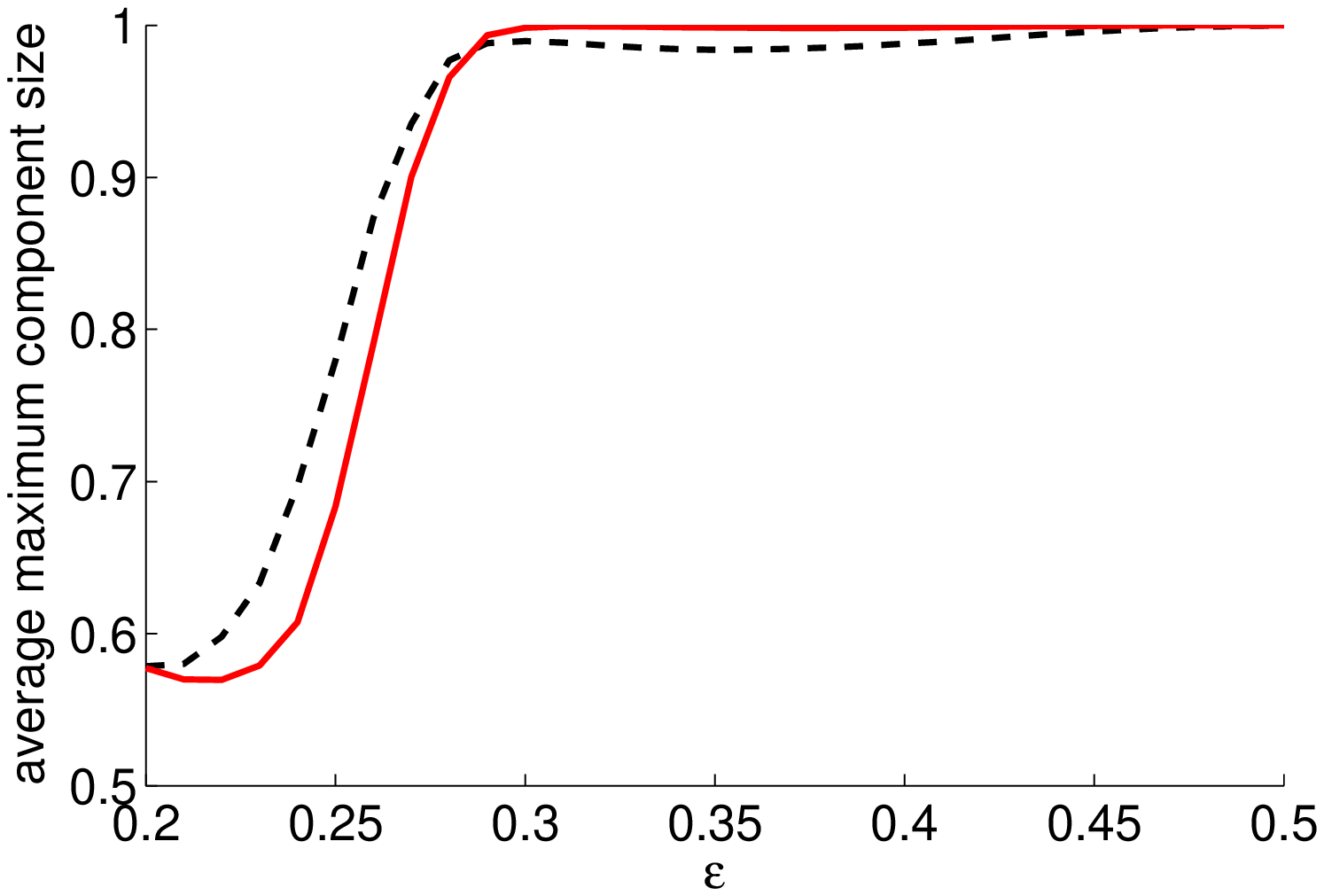}
    \label{subfig:compstat2}
  }
  \caption{Average maximum relative component
    size dependent on $\epsilon$ for 5000 runs, 500 agents and
    different values of $\mu$. The agents' initial behaviour is random
    according to a uniform distribution, the initial network is empty. For the
baseline model without group influence, a decrease of the convergence speed
$\mu$ leads to an increase of the maximum component size for all
$\varepsilon$. This
holds only for open-minded agents (high $\varepsilon$) in case of our model.}
  \label{fig:compstat}
\end{figure}

What is the effect of variations to the population size $n$ and the
convergence speed $\mu$ on our findings? To explain this we consider
how both parameters affect the consensus frequency in our model and the baseline
model.  In both cases, an increase in the population size usually leads to a
decreased consensus probability as it becomes more likely that a single agent
with extreme initial behaviour is separated from the rest of the
population. For open-minded agents, this effect is amplified by the the faster
dynamics towards the center in the local cultures model. Thus, we observe that
the reduction in frequency of consensus is greater than in the baseline
model
(cf. Figure \ref{fig:consensus_freq}).
With respect to the average maximum component size, an increase in the
population size increases the advantage of the local cultures model compared to
the baseline model for narrow-minded agents (cf. Figure \ref{fig:max_cluster}).
In this case, the increased number of agents leads to a higher probability for
bridging links between two almost seperated components. Hence, these components
more often coalesce and thereby increase the difference between the maximum
component size in our model and that in the baseline model as $n$ grows. This is
also indicated by the increased distance of the two models' respective variance
peak (cf. Figure \ref{fig:max_stdev_cluster}).
If we decrease the convergence speed $\mu$, we observe an increase in the
consensus frequency for both models for all thresholds $\varepsilon$ as the
agents' behaviour moves slower towards the center. With respect to the maximum
component size, this only holds for the baseline model. Figure
\ref{fig:compstat} shows that this quantity decreases for narrow-minded agents,
i.e. if $\varepsilon$ is small. The reason is that the coalescence of
components becomes less likely for smaller values of $\mu$ as interaction with
a bridge link between two almost seperated components becomes less effective in
this case.

\section{Discussion}
\label{sec:discussion}

The present paper set out to study the emergence of local cultures.  To
do so, it focused on the first stage of the process where agents need to
obtain consensus on acceptable business practice.  Within a
bounded-confidence model of opinion dynamics, we added a feedback
mechanism between a agent's behaviour and the evolving agent network.  The
effect of this mechanism depends on the value of the interaction
threshold $\varepsilon$.  In comparison with the baseline model
\citep{deffuant00}, our feedback increased the likelihood of consensus
for narrow minded agents (small $\varepsilon$) as the group effect may
foster a coalescence of otherwise separated components.  For open-minded
agents (large $\varepsilon$), the likelihood of consensus decreased
because the group effect worked to speed up convergence as compared to
the baseline case.  In some instances, this convergence was too fast for
all agents to reach consensus.  However, this constellation still had a
substantial proportion of component agents finding consensus (the maximum
component size was almost as large as in the baseline case).  The fact
that all component agents want to maintain their networks thus leads to
behavioural constraints that may impede full consensus for very
open-minded agents but increases it for narrow-minded ones.  As local
cultures can also emerge within a sub-population, the aforementioned
results suggest that the desire to maintain interaction networks has a
positive effect on the emergence of (full or partial) consensus, which
would then form the basis of a local culture.

A next step in advancing the model would consist in a benchmark against
data.  Unfortunately, the key model parameters (especially
open-mindedness of agents) are very difficult to operationalise and
thereby measure.  As a result, any data investigation would probably have
to rely on qualitative, case-study evidence investigating how component
agents choose to interact with each other and whether a concern for one's
past partners does exist.  Such findings would give an inclination of
whether the mechanisms proxied in the model are actually at work. Beyond
a data benchmark the link between open-mindedness and group effects on
consensus could be investigated experimentally. Participants could be
surveyed on open-mindedness and would be allocated to two groups
accordingly. The experiment could then study in how far the consensus
dynamics differ between both groups.

A second avenue for expanding the present paper consists in model
extensions.  Two aspects spring to mind.  First, one could investigate
the effect of heterogeneity among agents regarding their open-mindedness
($\varepsilon$).  Recent contributions \citep{lorenz08} suggest that
heterogeneity plays a substantial role for the likelihood of consensus in
the baseline model.  Second, one could introduce non-empty initial
networks (in-groups) to proxy that entrepreneurs in components often have
an initial set of acquaintances from living in the area or studying in
the same university.  Given the contrasting effect of groups on
consensus, it would be interesting to investigate in how far non-empty
initial in-groups affect it.  In a more general theoretic context, it
would pay to apply this model to the emergence of norms and conventions
in general.  Given the more expansive body of research in this field,
opportunities for benchmarking the model's results against other existing
studies will probably arise.  It would be particularly interesting to see
whether our counterintuitive result on group-effects and consensus is
applicable to other constellations.

\end{document}